\documentclass{aa}
\usepackage{booktabs} 
\usepackage{amsmath} 
\usepackage{graphicx}
\usepackage{amssymb}
\usepackage{amsmath}
\usepackage{float}
\usepackage{txfonts}
\usepackage{gensymb}
\usepackage{hyperref}
\usepackage{subfig}
\usepackage{threeparttable}
\usepackage{multicol}
\usepackage{multirow}
\usepackage{longtable}
\usepackage{rotating}
\usepackage{tabularx}
\usepackage{float}
\usepackage{afterpage}
\usepackage{placeins}
\usepackage[normalem]{ulem}
\usepackage[figuresright]{rotating}
\usepackage[dvipsnames]{xcolor}
\usepackage{placeins}
\usepackage{lscape}

\begin{document}
	
\authorrunning{Hu \& Soubiran}

\titlerunning{Metallicities of old open clusters}
	
\title{Metallicities of old open clusters: A new Galactic map}

\author{Qingshun Hu \inst{1,2} \and  Caroline Soubiran \inst{2}}
	
\institute{School of Physics and Astronomy, China West Normal University, No. 1 Shida Road, Nanchong 637002, People's Republic of China, (\email{qingshun0801@163.com}\label{inst1}) \and Laboratoire d'Astrophysique de Bordeaux, Univ. Bordeaux, CNRS, B18N, all\'ee Geoffroy Saint-Hilaire, 33615 Pessac, France (\email{qingshun.hu@u-bordeaux.fr, caroline.soubiran@u-bordeaux.fr}\label{inst2})}
	
\date{Received 26 March 2025 / Accepted 22 May 2025}
	
\abstract{Old open clusters (OCs) can constrain the chemical evolution of the Galactic disc through their metallicity gradients and age-metallicity relation but they are affected by low statistics.}
{This work aims to determine precise and homogeneous metallicities for a number of old clusters ($\geq$ 500 Myr) from all-sky catalogues of stellar parameters leveraging Gaia spectrophotometry. Our purpose was to revisit the metallicity distribution of the oldest OCs as a function of their Galactic position and age with improved statistics.}{Several catalogues of stellar parameters have been cross-matched to the most recent census of OCs and their members. The median metallicities per cluster and per catalogue were evaluated by comparison to high-resolution spectroscopy. The best performance is achieved when only bright giants are considered. Metallicity maps are presented and analysed, as well as trends of the distribution.}
{Our sample includes $\sim$600 old OCs with a typical precision of 0.05 dex in metallicity. We identified metal-poor or metal-rich clusters never studied before, as well as moving groups as the remnants of dissolving clusters. Galactic maps show a smooth decrease in metallicity from inside to outside the disc. Metal-rich and metal-poor clusters exist at all ages but dominate respectively in the inner and the outer disc, with different scale heights.  The radial metallicity gradient was found to have a knee shape with a steep value of $-$0.084~$\pm$~0.004~dex kpc$^{-1}$ in the inner side and $-$0.018~$\pm$~0.056~dex kpc$^{-1}$ outside the knee. The inner radial gradient flattens with age. Vertically, the metallicity gradient is $-$0.415~$\pm$~0.030~dex kpc$^{-1}$. The large scatter in the distribution of metallicity versus age is nicely explained by the superposition of OC populations standing at different galactocentric distances, each with its own mean metallicity and small dispersion, less than 0.08 dex in radius bins of 1 kpc.}
{Our results are consistent with a negative radial metallicity gradient of interstellar matter that was present in the disc when the clusters formed. The low metallicity dispersion in each radius bin reflects weak radial mixing. Our OC sample also indicates that most of the chemical enrichment of the Galactic disc occurred before they were formed.}

	
\maketitle


\section{Introduction}

For a long time, old open clusters (OCs) have been used as tracers of the structure and evolution of the Milky Way, as reviewed by \citet{Friel+1995}. After the pioneering work by \cite{Janes+1979}, \citet{Friel+2002} convincingly demonstrated that the metallicity of the old OCs is very valuable to probe the earliest stages of the Galactic disc through metallicity gradients and the age-metallicity relation (AMR). With their sample of 39 OCs older than 700 Myr they showed that the radial metallicity gradient seems to evolve with time, being steeper for the oldest clusters. This result gives strong constraints on the rates and histories of star formation, infall, and radial flows that are used in models of Galactic chemical evolution.  Metallicity gradients and their time evolution have been discussed in many papers, from the observational point of view with various tracers, or theoretically, for instance by \citet{Nordstr+2004, Netopil+2016, Anders+2017, Minchev+2018, Spina+2021b, Vickers+2021, Myers+2022, Magrini+2023, GaiaCollRecio-Blanco+2023, Joshi+2024, Haywood+2024, Carbajo-Hijarrubia+2024,Renaud+2025, Yang+2025}, to cite a few. OCs are among the best indicators of the trends in metallicity due to their wide spatial and age coverage, but studies based on old OCs have been limited until now by the small number of OCs with reliable metallicity.

The knowledge of the OC population has considerably improved in recent years thanks to the advent of Gaia astrometric and photometric data \citep{Gaia+2016}, as reviewed by \citet{Cantat-Gaudin+2024}. In particular, Gaia EDR3 \citep{Gaia+2021} and DR3 \citep{Gaia+2023} enabled the discovery of many new clusters and the improvement of membership determination, thanks to the precision, homogeneity, and all-sky nature of the Gaia data. An updated census of OCs, with a list of their members, is now available in the form of ready-to-use catalogues \citep[e.g.][]{Hunt+2023, Hunt+2024}, which are allowing new investigations of the global properties of the full population and their relation to the Milky Way structure and evolution. The parameters of OCs (e.g. distance, size, mass, extinction, metallicity, age) are usually determined via automated methods owing to the several thousands of objects involved \citep{Cantat-Gaudin+2020,Dias+2021,Hunt+2023,Cavallo+2024}. However, degeneracies between parameters, notably between extinction and metallicity, as well as priors or the non-representative training samples that are used in these methods, may lead to significant biases. Metallicities obtained by spectroscopy of individual members are expected to be more reliable.

Spectroscopic metallicities of OCs are still sparse and inhomogeneous despite significant efforts to determine cluster properties in recent years. The brightest OCs have been observed frequently at high spectroscopic resolution, but fewer than 100 objects are currently concerned. \citet{Netopil+2016} compiled high-quality spectroscopic metallicities from the literature for 88 OCs, 55 of which were older than 500 Myr. The OCCASO project \citep{Casamiquela+2016} intended to improve the situation by observing poorly studied OCs and added 36 clusters (30 older than 500 Myr)  to the list of OCs with a reliable metallicity deduced from high-resolution spectroscopy  \citep{Carbajo-Hijarrubia+2024}.

Multiplexed spectroscopic surveys with a typical resolving power of R$\sim$30\,000, such as the Gaia–ESO survey \citep{Randich+2022}, GALAH \citep{DeSilva+2015,Buder+2021} and APOGEE \citep{Abdurro'uf+2022}, dedicate a significant observing time to OCs, providing new metallicities for population studies. The forthcoming WEAVE \citep{Jin+2024}, 4MOST \citep{deJong+2019} and MOONS \citep{Cirasuolo+2014} will also target OCs. Spectroscopic surveys have already complemented Gaia with accurate radial velocities and chemical abundances of typically one million stars, and significantly enlarged the sample of clusters with well-defined properties. These excellent datasets have revitalised the studies of Galactic metallicity gradients as traced by OCs with a particular focus on their evolution over time \citep{Spina+2021b, Netopil+2022, Myers+2022, Spina+2022, Magrini+2023,Carbajo-Hijarrubia+2024,Palla+2024}. With a lower spectral resolution and a fainter limiting magnitude, LAMOST \citep{Wang+2022} provides metallicities for larger samples of clusters that were used to investigate the radial gradients  \citep{Fu+2022,Zhang+2024, Yang+2025}.

Although ground-based spectroscopic surveys have improved the number of clusters with a reliable estimate of metallicity, they are still missing many of them owing to their sky coverage, which is not complete. In addition, it is likely that the samples of OCs in spectroscopic surveys are strongly biased by selections due to observing strategies and by the limited number of known OCs before Gaia DR3. Gaia DR3 allowed a significant step forward in the determination of OC metallicities thanks to the all-sky survey of atmospheric parameters and chemical abundances of 5.6 millions stars based on the Radial Velocity Spectrometer \citep{Cropper+2018, Katz+2023, Recio-Blanco+2023}. \cite{GaiaCollRecio-Blanco+2023} analysed the metallicity [M/H] distribution of 503 OCs older than 100 Myr and highlighted the age dependence of the radial gradient. However, this sample was based on the list of known OCs prior to Gaia DR3 and was limited by the magnitude range of the Gaia spectroscopic survey. Gaia DR3 also provides accurate broad-band photometry for $\sim$1.5 billion sources, as well as low-resolution XP spectra for $\sim$219 million sources brighter than G=17.65 \citep{DeAngeli+2023, Montegriffo+2023}. A lot of effort has been dedicated in the last few years by several groups to inferring reliable metallicities of individual stars from this valuable (spectro)photometric dataset \citep{Anders+2022,Andrae+2023b,Zhang+2023,Khalatyan+2024,Fallows+2024,Ye+2025}. 
There is therefore a potential of determining the metallicity of many more OCs thanks to the new OC census and these all-sky catalogues of stellar parameters based on Gaia DR3.
The purpose of our work was to take advantage of these new data to determine the median metallicity of old OCs, with ages older than 500 Myr, by considering the individual metallicities of their members.  With improved statistics, we revisited the radial and vertical gradients and the age-metallicity relation (AMR) of OCs. First we assessed the precision and accuracy of the OC metallicities determined from Gaia (spectro)photometry with members of well-known clusters, and by comparison to mean metallicities derived from high-resolution spectroscopy.

\section{Data and methods}

\subsection{The old OC sample}
 
\begin{figure}[htbp]
	\centering
	\includegraphics[angle=0,width=88mm]{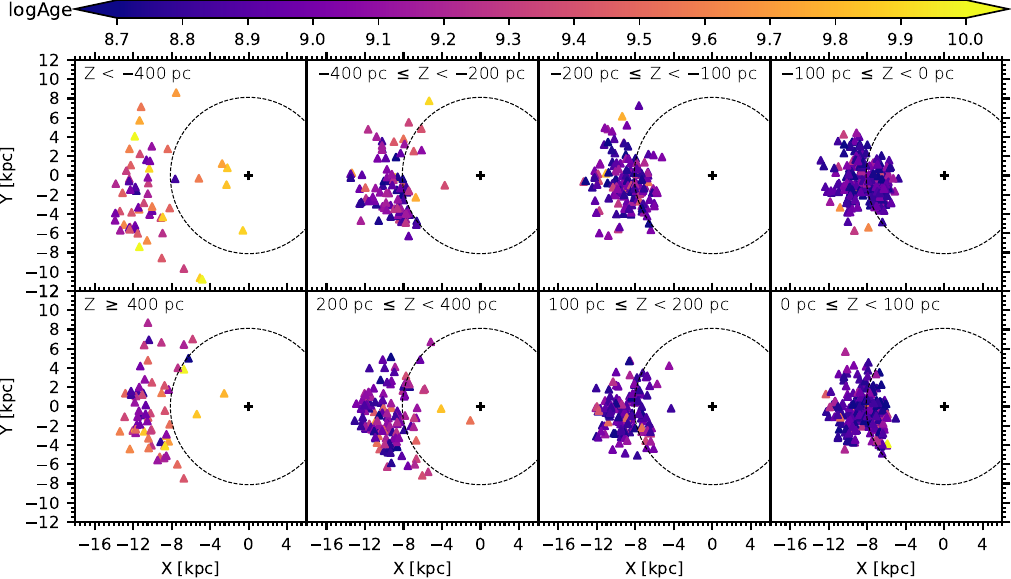}
	\caption{Age map on X-Y plane of 1,118 OCs older than $\sim$500 Myr (log(age)$\geq$8.7) taken from the gold sample of \citet{Cavallo+2024}, colour-coded by age from that catalogue, in different bins of distance to the Galactic plane Z. The Galactic centre is represented by a black cross. The solar circle is shown at R$_{\text{G}}$=8.122 kpc \citep{GRAVITY+2018}.}
	\label{fig: initial_sample}
\end{figure}

\citet{Hunt+2023} used Gaia DR3 data to build a homogeneous catalogue of more than 7,000 clusters, a large fraction unknown before, among which 4,105 are considered to be highly reliable clusters. The catalogue provides ages, extinctions, distances, as well as membership lists for all these clusters, and was updated by \cite{Hunt+2024}. \citet{Cavallo+2024} determined more robust parameters for most of these clusters using colour–magnitude diagrams combining photometric data from Gaia and 2MASS \citep{Cutri+2003}, and an artificial neural network trained on synthetic clusters. Ages, metallicities, extinctions, and distances were considered credible for $\sim$5,300 clusters in their gold sample, where isochrones showed a good agreement with the observational data. From this gold sample, we selected clusters older than $\sim$500 Myr (logAge50~$\geq$~8.7). We then removed globular clusters identified by \cite{Hunt+2024}, leaving us with 1,118 objects. We note that this selection includes 116 moving groups that we decided to keep since it is interesting to evaluate the consistency of metallicity among their members that would indicate that they are remnants of dissolved clusters. The distribution of this sample in Galactic coordinates is shown in Fig.~\ref{fig: initial_sample}, for different distance bins to the Galactic mid-plane. The values of galactocentric distance R$_{\text{G}}$ and of coordinates (X, Y, Z) are taken from \cite{Cavallo+2024} (no confidence interval is provided, the Sun is assumed to have R$_{\text{G}}$=8.122 kpc). We note that the highest clusters are essentially outside the solar circle and older on average than clusters close to the plane, in agreement with previous studies \citep[see e.g.][and references therein]{Cantat-Gaudin+2020}. It is well established that old clusters tend to be found at higher Galactic altitudes, mainly due to the vertical heating of the disc. The outer clusters might also show signs of the warp, as described by \cite{Cantat-Gaudin+2020}. Around the Galactic mid-plane, clusters younger than 1 Gyr largely dominate. This is consistent with the fact that open clusters with orbits maintaining them close to the disc tend to dissipate. The (X, Y) distribution of the lowest clusters is less extended than the higher ones, reflecting the detection limit due to the interstellar extinction at low Galactic latitudes, in particular at galactocentric distance smaller than 5.5 kpc.

The stars belonging to the 1,118 old OCs were then retrieved from the membership list of \cite{Cantat-Gaudin+2020} appended to that of \cite{Hunt+2024}, following \cite{Kos+2025}. As they did, we found both lists to be reliable but for some clusters one catalogue includes members that the other does not.

We tested the probability cuts on the list of \cite{Hunt+2024} (in \cite{Cantat-Gaudin+2020} there is already a cut at 70\%) without any improvement in the metallicity determination, likely because the cross-match to the catalogues of parameters (see below) implies a cut at G=17.65 which removes a good fraction of the low probability members. The resulting list includes 204\,024 stars, 2.5\% of which are reliable members from \cite{Cantat-Gaudin+2020}, but missing in \cite{Hunt+2024}.

The metallicities provided by \citet{Cavallo+2024} are not accurate enough to study the galactic gradients of old OCs. We noted some discrepancies with the literature that motivated this study with other sources of metallicities. As an example, we found the case of two clusters with metallicity values in \citet{Cavallo+2024} that are significantly different from the literature values. NGC~6791 is a populated and well-studied cluster, with a  very old age, and therefore an important target for Galactic studies. \citet{Cavallo+2024} give a metallicity of $-$0.4 while high-resolution spectroscopic studies give a high metallicity for it: \cite{Friel+2002} determined [Fe/H] in the range +0.13 to +0.17, the Gaia ESO survey gives [Fe/H]=+0.23$\pm$0.2 \citep{Randich+2022}, while \cite{Casamiquela+2021} found Fe/H]=+0.216$\pm$0.022. NGC~6583 is another case of low metallicity determined by \citet{Cavallo+2024}, [Fe/H]=$-$0.81, while the Gaia ESO survey gives [Fe/H]=+0.22$\pm$0.01 \citep{Randich+2022}.

\subsection{All-sky catalogues of metallicities}

In order to determine the metallicity of each OC in our sample, we gathered metallicities of individual members from all-sky catalogues that were built on Gaia DR3 photometry or on low-resolution BP/RP spectra \citep{Montegriffo+2023,DeAngeli+2023} that are available for $\sim$219 million stars, eventually combined to other multi-wavelength photometry. The first set of such metallicities is part of Gaia DR3 and includes 471 million stars \citep{Andrae+2023a}. It is not usable for our purpose, however, owing to substantial biases documented by the authors. Subsequently, a considerable effort was dedicated by different teams to improving the quality of the metallicities of individual stars determined from the Gaia data. In this work we considered the catalogues by \cite{Anders+2022,Andrae+2023b,Fallows+2024,Khalatyan+2024,Zhang+2023,Ye+2025}, that we cross-matched with our list of old OC members using Gaia DR3 source ID.  We applied the quality cuts recommended by the authors, and  our own restrictions. We kept only the members that are in the effective temperature range 4000 -- 6500 K, corresponding roughly to FGK-type stars which are known to have the most reliable metallicities for galactic studies \citep{Heiter+2014}. We also filtered out outlier members with metallicities lower than $-$1 or greater than 0.6. At the same time, we required at least three members to keep an OC in the sample. We give below a brief description of the methodology and input data that were used to build these six catalogues, and the result of the cross-match with the list of old OC members.

\subsubsection{\citet{Anders+2022}}\label{Anders22}

\citet{Anders+2022} determined the stellar parameters of 362 million stars, including metallicites [M/H] and extinctions Av, for stars brighter than G~$=$~18.5 in Gaia EDR3. As input data of the StarHorse code \citep{Queiroz+2018}, parallaxes and G, BP/RP magnitudes from Gaia were used, together with multi-wavelength photometry from Pan-STARRS1 \citep{Chambers+2016}, SkyMapper \citep{Onken+2019}, 2MASS \citep{Cutri+2003} and AllWISE \citep{Cutri+2013}. The method is based on isochrone-fitting (therefore sensitive to inaccuracies in the underlying stellar models) with priors on the metallicity, geometry, and age characteristics of the main Galactic components, and on interstellar extinction.

Together with our cut in effective temperature, we also applied the quality cut \verb|sh_outflag|  = `0000' recommended to filter out problematic results. The cross-match to the list of OC members gives 98,252 stars, after keeping only OCs with at least three members, allowing the determination of metallicity for 1,107 clusters, as reported in Table~\ref{t:summary_numbers}.

\subsubsection{\citet{Andrae+2023b}}

\citet{Andrae+2023b} utilised a machine learning algorithm to derive estimates of stellar metallicity [M/H] and other parameters for approximately 175 million stars from Gaia DR3 XP spectra, in combination with CatWISE magnitudes \citep{Marocco+2021} to reduce degeneracy between T$_{\text{eff}}$ and dust reddening. Parallaxes were also used to improve the accuracy of the estimates. Uncertainties on metallicities are not provided. The methodology is a  supervised learning approach with training on APOGEE data. For this study we decided to compare the results obtained with the full catalogue, and with the sample of over 17 million bright (G < 16) red giants defined by \citet{Andrae+2023b} who considered the [M/H] values of these stars to be the most precise and pure. We applied our cut in T$_{\text{eff}}$ to both samples, even if the sample of giants is already limited to T$_{\text{eff}} <$ 5200 K.

We obtained metallicity data for 81\,384 members in 1\,113 old OCs when the full catalogue is considered, and 7,994 members in 597 old OCs for the sample of giants, as reported in Table~\ref{t:summary_numbers}.

\subsubsection{\citet{Zhang+2023}}

\citet{Zhang+2023} employed a forward model to estimate stellar atmospheric parameters, including metallicity [Fe/H], for $\sim$220 million stars based on their Gaia DR3 XP spectra combined to 2MASS and WISE photometry. The metallicity of the stars was determined using a data-driven approach that leverages a neural network trained on spectroscopic data from the LAMOST survey, which provides broad coverage of different spectral types.

Together with the T$_{\text{eff}}$ cut, we also applied two recommended quality cuts (\verb|quality_flags| < 8 and \verb|feh_confidence| > 0.5 ).
The cross-match to the list of OC members gives 41,462 stars, in 941 clusters, as reported in Table~\ref{t:summary_numbers}.

\subsubsection{\citet{Khalatyan+2024}}

\citet{Khalatyan+2024} presents the SHBoost catalogue with stellar parameters, including metallicity [M/H] and extinction, for over 217 million stars. The method is a supervised machine learning algorithm applied on Gaia DR3 XP spectra complemented by Gaia astrometry and broad-band photometry from Gaia, 2MASS and AllWISE. The main difference with the catalogue of \citet{Andrae+2023b} based on a similar technique is the use of more extended training sets which includes all the spectroscopic surveys, Gaia-ESO, APOGEE, LAMOST, Radial Velocity Experiment \citep[RAVE,][]{Steinmetz+2006, Steinmetz+2020}, GALAH \citep{DeSilva+2015}, and Sloan Extension for Galactic Understanding and Exploration \citep[SEGUE,][]{Yanny+2009}, and specific sets of stars poorly represented in these surveys.

With a quality cut \verb|xgb_met_outputflag| = '0', which means the standard deviation of metallicity is less than 0.3, combined to our restriction of the T$_{\text{eff}}$ range and the number of members per cluster, we ended up with 73,418 stars in 1,093 OCs, as reported in Table~\ref{t:summary_numbers}.

\subsubsection{\citet{Fallows+2024}}

\citet{Fallows+2024} estimated atmospheric parameters and abundances, including [Fe/H], from Gaia BP/RP spectra, for over 182 million stars. The method is an uncertain neural network trained on APOGEE, that also uses broad-band photometry from Gaia DR3, 2MASS and WISE.

We obtained metallicity data for 1,107 OCs, corresponding to 78,996 members, as reported in Table~\ref{t:summary_numbers}.

\subsubsection{\citet{Ye+2025}}\label{Ye24}

\citet{Ye+2025} provide atmospheric parameters, including [M/H], for 68 million stars by fitting Gaia DR3 XP spectra with synthetic spectra based on model atmospheres. A neural network was trained on APOGEE data to predict flux corrections as a function of wavelength for each target star. Corrections for systematic errors in the XP spectra related to colours, magnitudes, and extinction, lead to improved precision in the relative spectrophotometry of the Gaia XP data. However, since the results are provided only for stars that meet some criteria on the flux correction, the catalogue is much smaller than the initial sample of $\sim$219 million XP spectra, missing in particular many objects in the Galactic plane due to extinction. We obtained metallicity data for 363 OCs, corresponding to 13,511 members, as reported in Table~\ref{t:summary_numbers}.

\subsection{Metallicity per cluster and per catalogue}

For each catalogue we simply adopted the median metallicity of the member stars of each cluster as the metallicity of the cluster, with the median absolute deviation (MAD) as its uncertainty. 

\begin{table}
\centering
\caption{Summary of the cross-match between the old OC sample and the metallicity catalogues.}\label{t:summary_numbers}
	\begin{tabular}{l | r | r | r}
		\hline\noalign{\smallskip}
		\hline\noalign{\smallskip}
Xmatched catalogue &  Old OC & members & MAD \\
\hline\noalign{\smallskip}
\citet{Anders+2022} & 1,107 & 98,252 & 0.11 \\
\citet{Andrae+2023b} full & 1,113 & 81,384 & 0.11 \\
\citet{Andrae+2023b} giants & 597 & 7,994 & 0.05 \\
\citet{Zhang+2023} & 941 & 41,462 & 0.10\\
\citet{Khalatyan+2024} & 1,093 & 73,418 & 0.06 \\
\citet{Fallows+2024} & 1,107 & 78,996 & 0.11 \\
\citet{Ye+2025} & 363 & 13,511 & 0.13 \\
\hline\noalign{\smallskip}

\end{tabular}
\centering
\tablefoot{The number of clusters and cluster members retrieved in each catalogue is provided together with the typical value of MAD computed with the median metallicity per cluster.}
\end{table}

Table~\ref{t:summary_numbers} summarises the number of clusters for which a median metallicity was computed, with the number of members that were retrieved. The catalogue of \citet{Anders+2022}
is the only one that used broad-band photometry alone. Therefore it is not affected by the limiting magnitude of the XP spectra (G $<$ 17.65) and this is why more members are retrieved. However this does not translate into a much larger number of old OCs compared to the other catalogues. \citet{Andrae+2023b}, \citet{Zhang+2023}, \citet{Khalatyan+2024}, and \citet{Fallows+2024} also provide metallicities for about 1000 old OCs. The cross-match of our list of cluster members to the vetted subsample of bright giants from \citet{Andrae+2023b} clearly gave fewer OCs. \citet{Ye+2025} made stringent quality cuts resulting in the smallest sample of old OCs. 
The typical uncertainty listed in Table~\ref{t:summary_numbers} for each catalogue gives an idea of their precision. The best precision is achieved by \citet{Andrae+2023b} due to bright giants being very reliable targets for abundance determinations, but \citet{Khalatyan+2024} also seems to be good for the precision. Metallicities based on the catalogue of \citet{Ye+2025} exhibit the lowest precision. Since it also gives the smallest sample of clusters, we do not consider it in the following.

We provide a catalogue of 1\,113 old OCs with up to six values of metallicity computed from the stellar parameters determined by  \citet{Anders+2022}, \citet{Andrae+2023b}, \citet{Zhang+2023}, \citet{Khalatyan+2024}, and \citet{Fallows+2024}, with two datasets from \cite{Andrae+2023b}, a full dataset and one with only bright giants. An excerpt of our OC catalogue, distributed by the CDS, is given in Table \ref{table:metallicity}.
For each of the six considered datasets of stellar parameters, we show in Figs~\ref{fig: HSC~1725}, \ref{fig: NGC~6791} and \ref{fig: Be73} the metallicity distribution of three representative clusters, HSC~1725, NGC~6791 and Berkeley 73 as a function of effective temperature and Gaia magnitude. NGC~6791 is a populated and well-studied cluster, metal rich and very old as mentioned previously. On the contrary, HSC~1725 is a newly discovered, younger ($\sim$600 Myr) and sparse cluster, according to \citet{Hunt+2024} and \citet{Cavallo+2024}. Berkeley 73 is another known metal-poor cluster, part of several spectroscopic surveys. The high metallicity of NGC~6791 is confirmed by the six catalogues. The median metallicities from \citet{Anders+2022} and \citet{Khalatyan+2024} and the full catalogue of \citet{Andrae+2023b} are close to zero when all the members are considered, but stars cooler than $\sim$5000 K, or brighter than $\sim$16, clearly show a higher metallicity that is consistent with the three other catalogues. Systematic effects are seen for the faintest stars. HSC~1725 is found to be metal-poor by all the catalogues, with median metallicities ranging from $-$0.34 to $-$0.75. The six catalogues provide consistent metallicities for Berkeley 73, ranging from $-$0.40 to $-$0.53.

\subsection{Comparison with high-resolution spectroscopy}
\label{s:comp_hr}

\begin{figure*}[htbp]
	\centering
        \includegraphics[width=78mm]{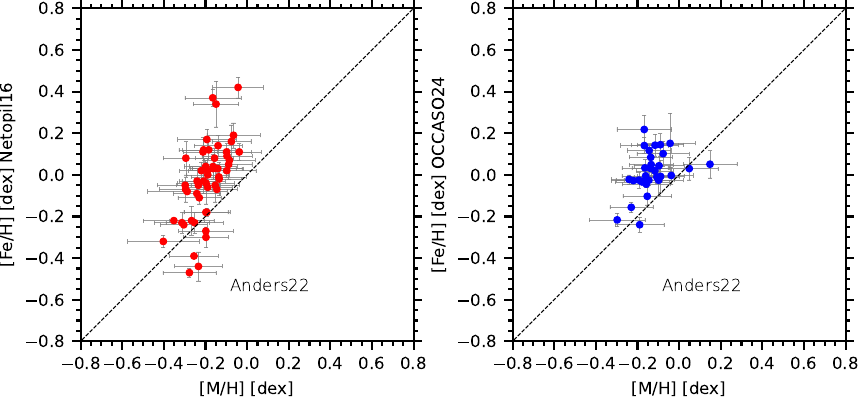}
        \includegraphics[width=78mm]{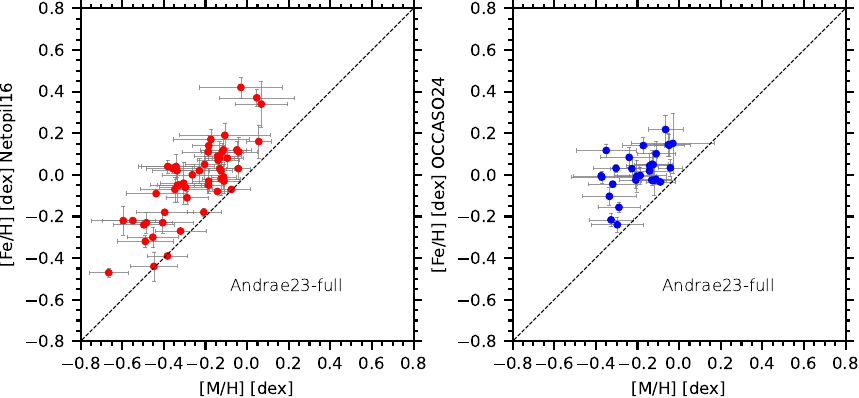}
        \includegraphics[width=78mm]{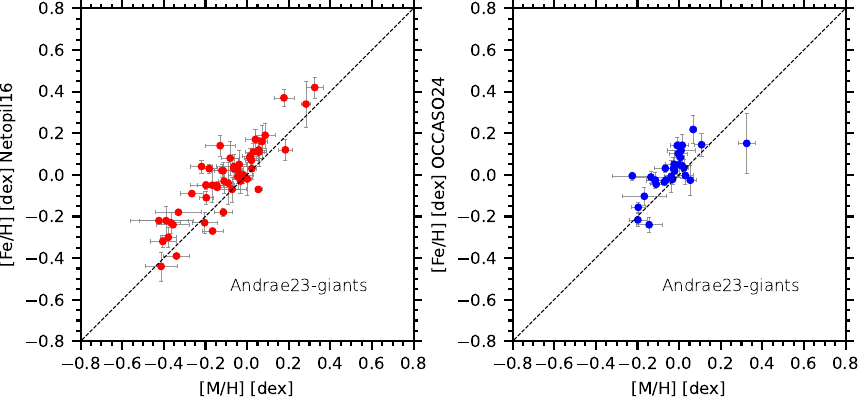}
		\includegraphics[width=78mm]{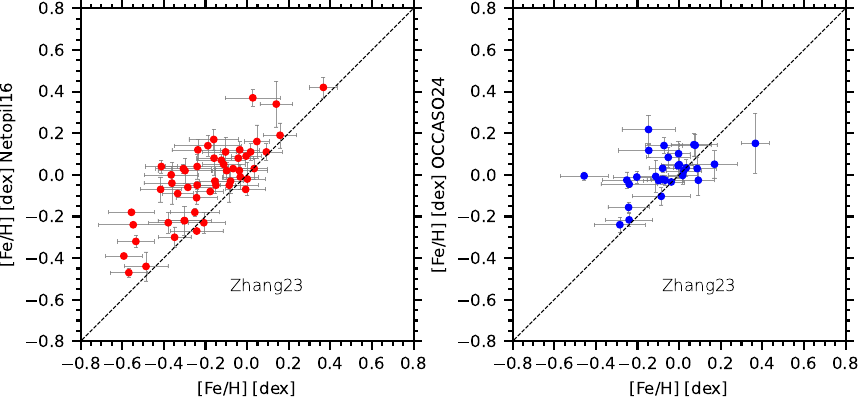}
        \includegraphics[width=78mm]{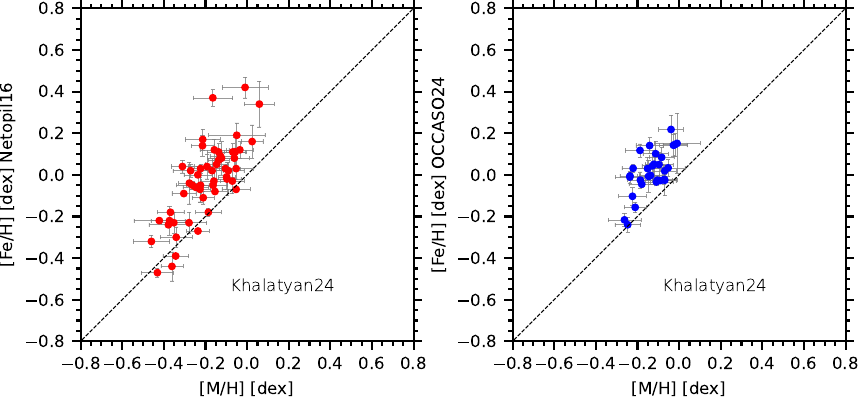}
        \includegraphics[width=78mm]{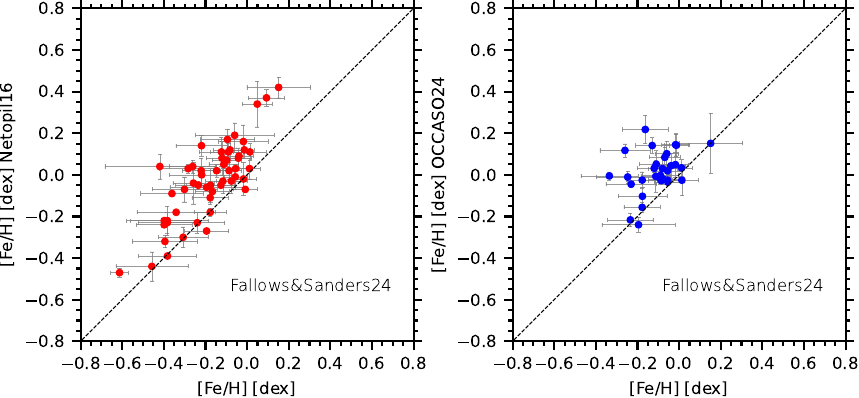}
	\caption{Comparison of calculated metallicities ([Fe/H] or [M/H]) from the six catalogues to high-resolution spectroscopic [Fe/H] from \citet{Netopil+2016} (red) and from \citet{Carbajo-Hijarrubia+2024} (OCCASO, blue). }
	\label{fig: Comparison_MFeh_literature}
\end{figure*}

In order to assess the reliability of the median metallicity values per cluster and per catalogue obtained above, we compared them to the metallicities of clusters obtained from high-resolution spectroscopy. We adopted as reference the high-quality samples from \citet{Netopil+2016} and from OCCASO \citep{Carbajo-Hijarrubia+2024}. The Comparisons are shown in  Fig.~\ref{fig: Comparison_MFeh_literature} for the six catalogues and the two reference samples. 

From these plots we can evaluate whether a given catalogue is able to retrieve the full range of metallicities that is present among old OCs of the reference samples, in particular of the sample of \citet{Netopil+2016} which extends from [Fe/H]$\sim$$-$0.5 to [Fe/H]$\sim$+0.4. The six catalogues performed reasonably well to retrieve the expected metallicity of metal-poor clusters ([Fe/H]$<$0). However, they significantly underestimated the metal-rich metallicities ([Fe/H]$>$0), except \cite{Zhang+2023} and \cite{Andrae+2023b} limited to bright giants. The metallicities from \citet{Anders+2022} occupy a particularly narrow range that does not reflect the variety of chemical compositions deduced from high-resolution spectroscopy. To lesser extent, metallicities from \cite{Khalatyan+2024} also show a similar trend, documented by the authors, which they were able to reduce with an appropriate calibration. It is worth noting that the six catalogues exhibit an offset, the clusters appear in general to be more metal-poor than in the reference samples.

The smallest offset is obtained with the sample of bright giants from \citet{Andrae+2023b}. The difference in zero-point of metallicities does not affect the value of gradients that is computed in the next section. Finally, the dispersion of the comparisons gives an idea of the precision of each catalogue. The lowest dispersion is seen for the sample of bright giants from \citet{Andrae+2023b}, while the five other catalogues exhibit a significant dispersion. From these comparisons to high-resolution spectroscopy, we conclude that the sample of bright giants from \citet{Andrae+2023b} gives the best performance in retrieving the full range of metallicities of old OCs, with the best precision and accuracy, at the cost of losing nearly half of the old OCs initially selected.

\section{Peculiar old open clusters}
\label{s:peculiar_oc}

We mention here a few old OCs that are remarkable for their extreme metallicity values or inner disc positions which make them interesting for further investigation. They are also important owing to their weight in the determination of the metallicity gradients. We also investigated the properties of some old moving groups that might be OC remnants.

\subsection{Metal-poor clusters}

Here we consider the clusters with the most reliable metallicity, relying on at least ten members in the full catalogue of \cite{Andrae+2023b}. A few of them with [M/H]$<$ $-$0.6 have already been identified as metal-poor clusters and have sometimes been observed in spectroscopy (e.g. Berkeley 29, Berkeley 31, NGC 2324, FSR~0975,..), but for most of them we provide the first determination of metallicity. For example, HSC~1794 and HSC~1835, discovered by \citet{Hunt+2023}, are new old OCs and are interesting objects. We determined their metallicities for the first time to be around $-$0.70, based on 12 members in both cases, with a MAD of 0.09 and 0.15 respectively. In addition, there are known OCs that did not yet have a determination of metallicity, such as Auner~1, CWNU~1714, UBC~626, UBC~1326, MWSC~1226 and UBC~1318. Their metallicity in this work is found to be around $-$0.6 but should be redetermined, owing to the offset between \cite{Andrae+2023b} and high-resolution spectroscopy reported in the previous section. MWSC~1226 is of particular interest because it has 13 bright giants that confirm its metal-poor nature. Tombaugh~2 is also confirmed to be among the most metal-poor old OCs thanks to 39 bright giants from \cite{Andrae+2023b}. HSC 1293 is another good case of a metal-poor cluster with ten giants.

The 90 metal-poor clusters with [M/H]$<$ $-$0.5 span the full age range and distance to the Galactic mid-plane. However all but three of them (Theia 8068, UBC 504, NGC 3255), lie in the outer disc (R$_{\text{G}}> 9$ kpc). The farthest OC is Berkeley 29, which was found to have [M/H]$=-0.665\pm$0.09, based on 19 members.
The three oldest clusters of the sample (Gaia 2, ESO 92-18, ESO 429-05) with an age of $\sim$10 Gyr according to \cite{Cavallo+2024}, all lie at large distance from the Galactic plane (Z$<-$500 pc), and have a metallicity of [M/H]$\sim-$0.50. \cite{Garro+2023} considered Gaia 2 as a globular cluster and ESO 92-18 as an OC. 

It is worth noting that only one cluster with [M/H]$<-$0.6, Theia 4338 (CWNU 1378), remains in the OC sample built from the giants of \cite{Andrae+2023b}.

\subsection{Metal-rich clusters}
\label{s:MR_OC}

Here we consider the sample of clusters with metallicity deduced from the bright giants of \cite{Andrae+2023b} which performs better in this metallicity range, as shown in the previous section. In this sample, NGC~6791 is an outlier with a metallicity well above all the other clusters, with [M/H]=+0.325$\pm$0.04, based on 109 giants. NGC~6791 is also exceptional because of its age of $\sim$10 Gyr and its position in the inner disc and large distance to the galactic mid-plane (Z=689 pc). NGC~6253 and Ruprecht~134, also known to be metal rich, very old and in the inner disc, are confirmed to have a high metallicity, greater than +0.25 with an uncertainty lower than 0.04, based on 53 and 12 giants, respectively. In addition to these well-known OCs, there are several OCs whose metallicities have been poorly or never studied, such as FoF~861, Juchert~13, UBC~537, UBC~268, UBC~321, and UBC~1031. Their metallicity around [M/H]=+0.2 is determined for the first time in our work with an uncertainty lower than 0.05.

We note that the 41 old OCs with [M/H]$\geq$ +0.15 all lie within the solar radius at R$_{\text{G}}<7.7$ kpc. Of these, HSC 2966 is the innermost at R$_{\text{G}}$=4.4 kpc. 

\subsection{Inner-disc clusters}
\label{s:inner_oc}
We count 11 clusters with R$_{\text{G}}<5.5$ kpc and a metallicity determined from the full catalogue of \cite{Andrae+2023b}. They all have a high extinction according to \cite{Cavallo+2024}, from Av=1 up to Av=5.3 for HSC 2966. All but one (Theia 1813) have recently been discovered by \cite{Hunt+2023} and six of them are classified as moving groups by \cite{Hunt+2024}. If we focus on the best metallicities relying on giants, 
we note the super metal-rich HSC 2966, already mentioned in Sect. \ref{s:MR_OC}. HSC 166, HSC 2878 and HSC 2867 have [M/H]$\sim$+0.1 and stand at R$_{\text{G}}<4$ kpc. Two other clusters, HSC 113 and HSC 172, appear as outliers with their low metallicity, respectively $-$0.48 and $-$0.40, and their high Z. Inner clusters are rare and very important for investigate the influence of the bar in the disc \citep[e.g.][]{Haywood+2024}. These newly discovered clusters deserve further investigation to consolidate their properties, notably because of their high level of extinction. If their orbits can also be determined they will become valuable probes of the inner disc.

\subsection{Moving groups}

\citet{Hunt+2024} classified clusters into bound OCs and unbound moving groups based on their mass and radius. If the members of a moving group are coeval stars, one would expect to find similar metallicities for them. We examine here the properties of the 116 moving groups in \cite{Andrae+2023b}. Moving groups are sparse and their metallicity mostly relies on  few members. Nevertheless some of them have a significant number of members together with a low metallicity dispersion that reflects a consistent [M/H] among the members. We note HSC~969 which has 23 members in \cite{Andrae+2023b}, including three giants giving [M/H]=$-$0.24$\pm$0.01, and HSC~962 with 21 members and also three giants giving [M/H]=$-$0.09$\pm$0.07. Both are close to the Galactic mid-plane with an age of about 1 Gyr according to  \cite{Cavallo+2024}. The most metal-poor moving group is HSC~1398, with [M/H]=$-$0.56$\pm$0.09 based on four giants. The most populated moving group is Theia 89 with 91 members in \cite{Andrae+2023b} including three giants giving [M/H]=+0.02$\pm$0.01. The low scatter in the metallicity of the members suggests that these objects are likely dissolving clusters. They are therefore useful probes of the early stages of star formation in the Galactic disc. 

Other moving groups are of particular interest due to their Galactic position. HSC~172 lies in the inner disc as mentioned previously, at 2 kpc above the Galactic plane. It has 23 members in \cite{Andrae+2023b} including seven giants giving [M/H]=$-$0.40$\pm$0.11, for an age of 6 Gyr. At 1 kpc below the Galactic mid-plane and close to the Galactic centre (1.4 kpc), HSC~113 is another very old moving group (7 Gyr), which has 11 members in \cite{Andrae+2023b} including four giants giving [M/H]=$-$0.48$\pm$0.24. HSC~172 and HSC~113 are mentioned below as outliers, owing to their extreme position and metallicity.  We also mention HSC~1293 with an age of 2.9 Gyr, Z=$-$1 kpc, R$_{\text{G}}$=12.4 kpc, and [M/H]=$-$0.51$\pm$0.14 relying on ten giants, and HSC~2867 with an age of 2.4 Gyr, Z=$-$485pc, R$_{\text{G}}$=2 kpc, and [M/H]=+0.10$\pm$0.06 relying on three giants.

Although the stars in these moving groups have a consistent metallicity suggesting that  they are of common origin, a further investigation is needed, in particular of their orbits, to understand why they have not completely dissolved under the effect of the Milky Way’s potential after such a long time.

\section{Probing the Galactic disc with old OCs}

\begin{figure*}[!ht]
	\centering
 \includegraphics[angle=0,width=146mm]{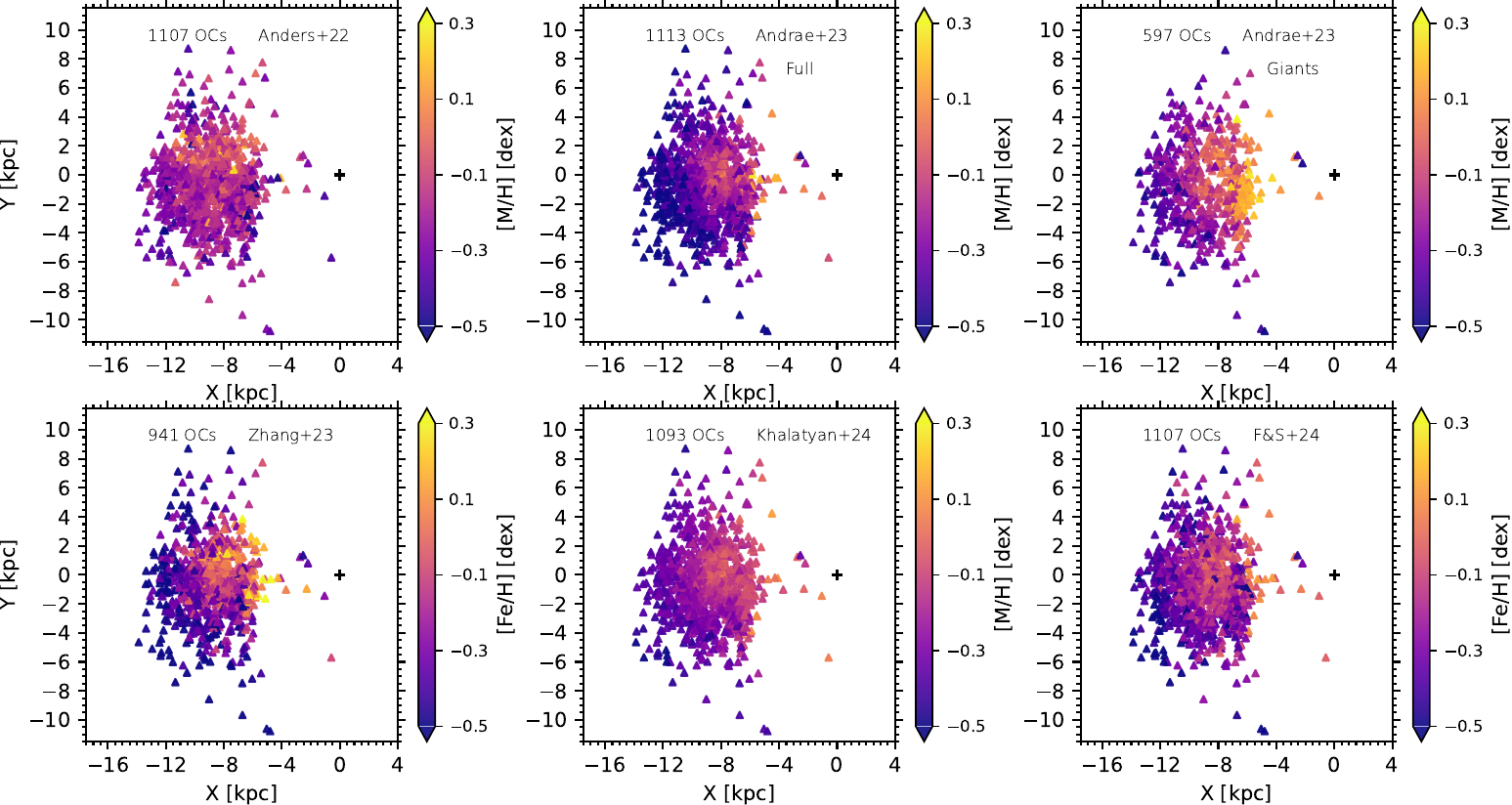}
	\caption{Metallicity maps on X-Y plane of OCs from six catalogues. The black plus sign in each panel marks the location of the Galactic centre.}
	\label{fig:XYZ_Feh_map}
\end{figure*}

\subsection{Metallicity maps}

Figure ~\ref{fig:XYZ_Feh_map} presents the metallicity distribution of old OCs on the X-Y projection plane of the Milky Way, as obtained from the six catalogues. Depending on the catalogue, we see more or less clearly a smooth evolution of the metallicity with galactocentric distance. All panels but that of \cite{Anders+2022} show globally the metal-rich clusters in the inner disc and the most metal-poor clusters distributed along an external half-annulus. In between, around the solar position, the old OCs have intermediate metallicities. The six maps also reflect the biases in metallicities that were already highlighted in the comparisons to high-resolution spectroscopy (Sect. \ref{s:comp_hr}). The above-noted offset translates here into an excess of metal-poor clusters, in particular with the full catalogue of \cite{Andrae+2023b}, and those of \cite{Zhang+2023} and \cite{Fallows+2024}, and into a lack of metal-rich clusters, notably with the catalogue of \cite{Khalatyan+2024}. The map traced by the clusters with metallicities retrieved from the bright giants of \cite{Andrae+2023b} is the one that is the most consistent with the maps presented by \citet{Eilers+2022} and \citet{Haywood+2024}, obtained with field red giant stars from APOGEE. In this map the metallicty of old OCs regularly decreases from the inner to the external parts of the disc. Because of the good agreement previously found in comparison to high-resolution spectroscopy, this is a new confirmation that the sample of bright giants from \cite{Andrae+2023b} provides realistic estimates of metallicity for old OCs. Therefore, for the following, we investigated the metallicity distribution using only the sample of 597 clusters with [M/H] derived from that catalogue. Although this sample includes fewer clusters than the other, the metallicities are of better quality and reliable enough to investigate Galactic gradients. The bulk of this sample (98\%) spans galactocentric radii between 5.5 and 15 kpc, up to 1.2 kpc in distance perpendicular to the mid-plane with a coverage in metallicity from about $-$0.55 to +0.25 dex. The few stars outside these boundaries are discussed in Sect. \ref{s:peculiar_oc}.

\begin{figure}[htbp]
	\centering
 \includegraphics[angle=0,width=88mm]{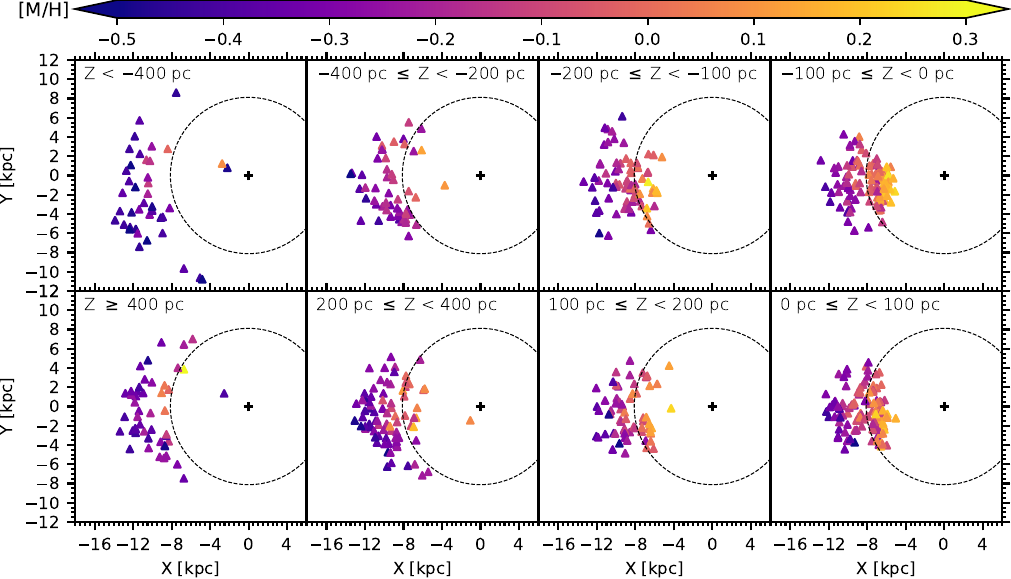}
	\caption{Metallicity maps on X-Y plane of the 597 OCs at different distances Z from the Galactic mid-plane. The location of the Galactic centre is marked, as is the solar circle.}
\label{fig: XYZ_An23_map}
\end{figure}

\begin{figure}[htbp]
	\centering
 \includegraphics[angle=0,width=76mm]{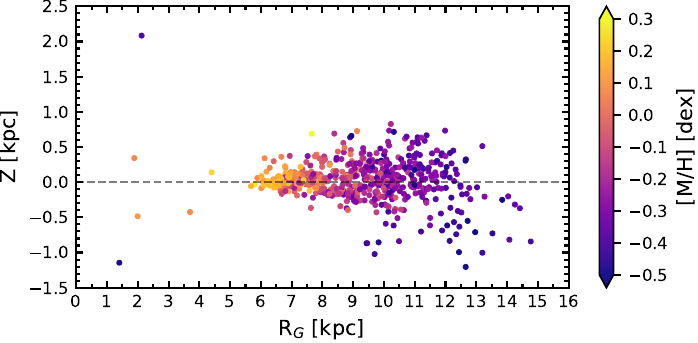} 	
 \caption{Height from the Galactic mid-plane versus distance from the Galactic centre for the 597 OCs colour-coded by their metallicities.}
	\label{fig:RG_Z_M}
\end{figure}

Similarly to Fig.~\ref{fig: initial_sample} where we showed the age maps in different bins of Z, we now show the metallicity maps of our sample of 597 OCs in the same way in Fig.~\ref{fig: XYZ_An23_map}. The metal-rich clusters are essentially close to the plane (only five OCs with [M/H]$>0$ at |Z|$>$400 pc), while the metal-poor ones appear in all Z bins. The smooth decrease in the metallicity with increasing radius is clearly seen at low |Z|. This evolution tends to shallow at higher |Z| where the metal-poor clusters start to dominate, inner clusters being less represented. Figure~\ref{fig: XYZ_An23_map} does not exhibit any significant variations depending on Y. Another representation of the OC distribution is given in  Fig.~\ref{fig:RG_Z_M}, with |Z| versus the galactocentric radius R$_{\text{G}}$ colour-coded by metallicity. The radial metallicity gradient is clearly visible, together with the scale height increasing with R$_{\text{G}}$. The distribution is not symmetric with respect to the mid-plane, there are more high-|Z| clusters in the Southern hemisphere. Together with the observation in Fig.~\ref{fig: initial_sample} that old and young OCs dominate respectively at high and low |Z|, these figures show that the metallicity distribution of the OCs is intrinsically related to their age and their Galactic position. 

\subsection{Metallicity gradients}

\begin{figure*}[htbp]
	\centering
 \includegraphics[angle=0,width=76mm]{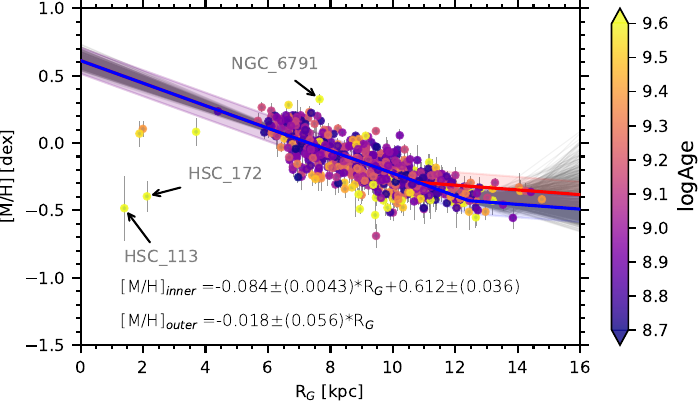}
 \includegraphics[angle=0,width=76mm]{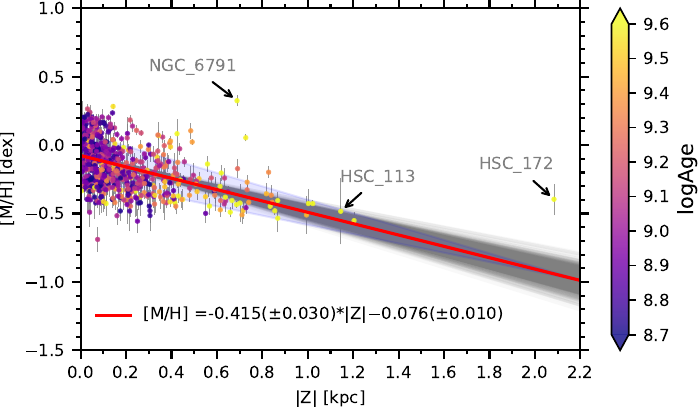}
  \includegraphics[angle=0,width=76mm]{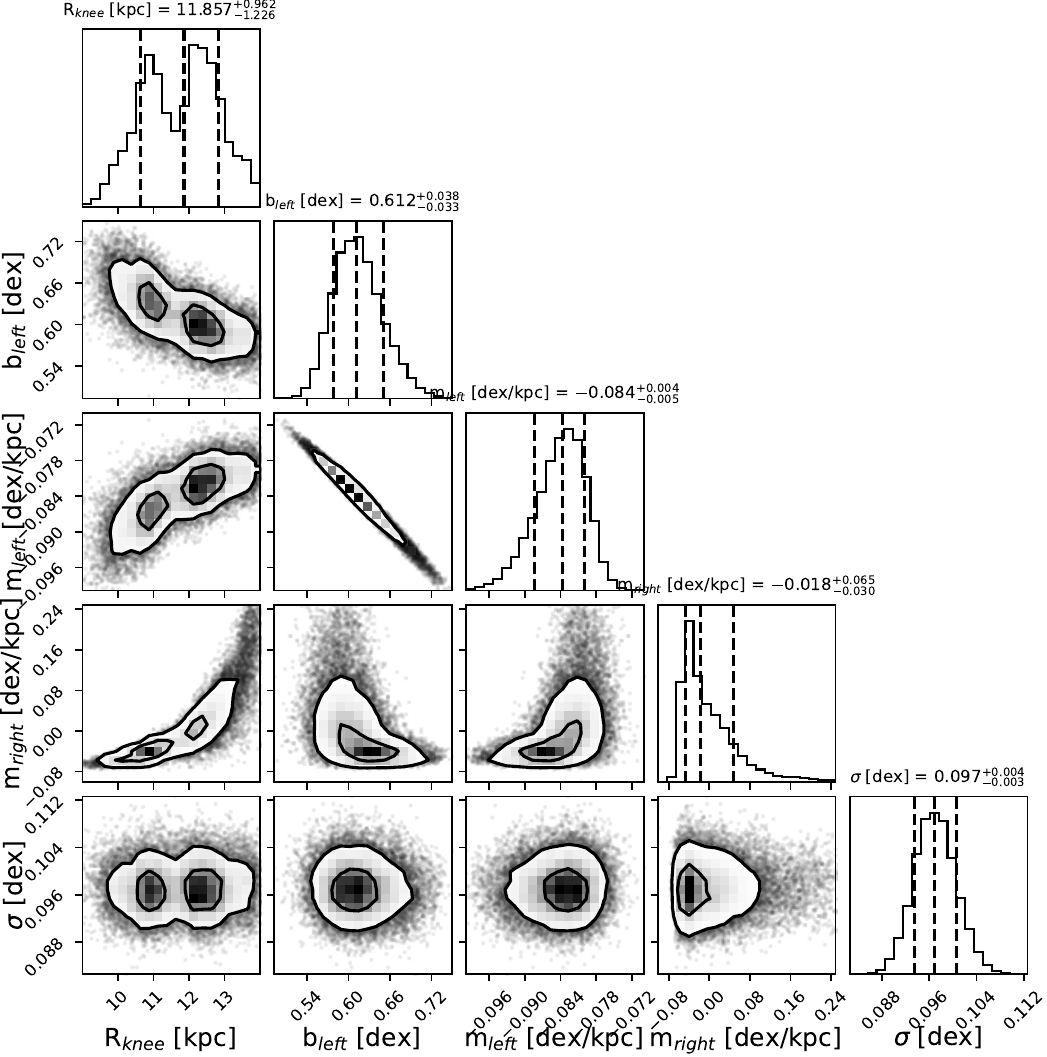}
 \includegraphics[angle=0,width=76mm]{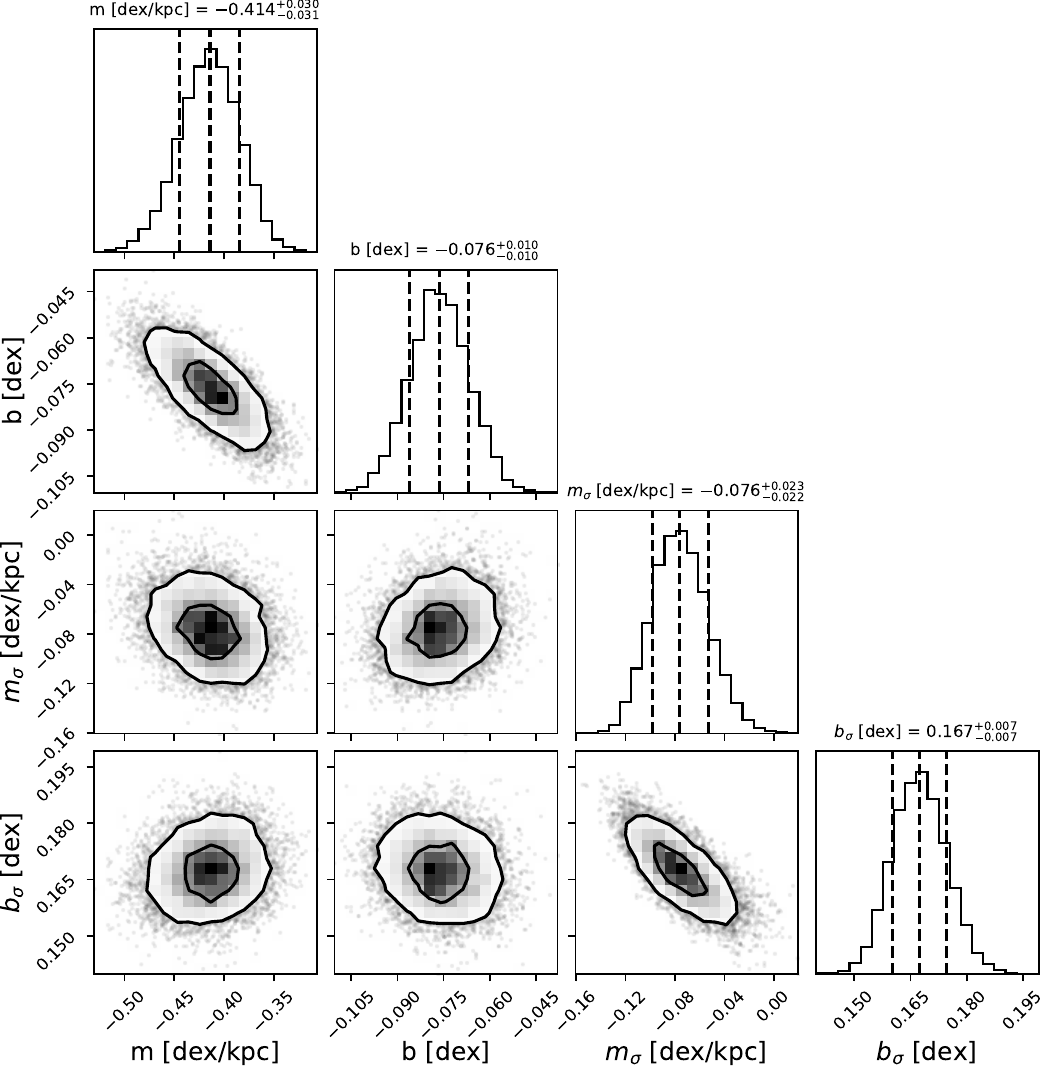}
	\caption{Top panels: Radial (top left) and vertical (top right) metallicity distributions of our sample of 597 OCs, colour-coded by age. The 1000 MC samplings (thin grey lines) in each panel are superimposed; the red and blue solid lines are the best fitting lines. The red or blue shaded bands in each panel corresponds to the 1$\sigma$ dispersion around the gradient. The knee positions of red and blue breaks in the top left panel are 10.80$\pm$0.04 kpc and 12.43$\pm$0.04 kpc, respectively. The three outliers are indicated. Bottom panels: Posterior distributions of the fit parameters of radial (bottom left) and vertical (bottom right) metallicity distributions.}
	\label{fig:Z_MFeh_relationships}
\end{figure*}

The maps presented above show that the metallicity of our 597 OCs evolves smoothly with the galactocentric radius R$_{\text{G}}$, and with the distance to the Galactic plane, Z. The metallicity distribution is plotted against R$_{\text{G}}$ and |Z| in Fig.~\ref{fig:Z_MFeh_relationships}; the colour indicates the age of the clusters. The vertical distribution is well-defined up to 1.2 kpc, one single OC (HSC 172) is above that value at 2 kpc. The radial distribution is well defined between 5.5 and 15 kpc; six OCs closer to the Galactic centre as mentioned in Sect. \ref{s:inner_oc}. Two metal-poor clusters in particular appear as outliers (HSC 172, HSC 113), their low metallicity placing them well below the expected value at this small radius. 
NGC 6791 is also an outlier in both distributions, being the most metal-rich cluster with an unusually large distance to the Galactic mid-plane. NGC 6791 is known for its eccentric orbit compared to typical disc objects \citep{Tarricq+2021}. These three outliers are very old clusters, and since HSC 172 and HSC 113 are newly discovered moving groups with extreme properties, they deserve an in-depth investigation to consolidate their properties, which is beyond the scope of this paper.

We quantified the radial and vertical metallicity gradients by fitting a linear relation between [M/H] and R$_{\text{G}}$, and between [M/H] and the absolute height |Z|. However, since the dispersions of the two distributions are higher than the uncertainties, we adopted the Bayesian linear gradient + scatter model, following \citet{Anders+2017}. For the radial gradient, we adopted a segmented linear function with a
shared knee point to take into account the flattening of the gradient that is visible in our data at large R$_{\text{G}}$. This change in slope in the radial metallicity gradient has been reported in many studies based on OCs and seems well established \citep{Carrera+2019, Donor+2020, Myers+2022, Magrini+2023, Spina+2022, Netopil+2022, GaiaCollRecio-Blanco+2023, Carbajo-Hijarrubia+2024}. However, the exact position of the break between 11 and 14 kpc is still subject to controversy. The details of the computation are provided in Appendix \ref{appendix_segmented}. The vertical gradient is mainly constrained by very old clusters at |Z|$>$ 500 pc, the dispersion being large at low |Z|. Within 200 pc of the mid-plane, the metallicity values of the clusters span the full range from -0.55 to +0.25. To take this into account we adopted a variable scatter in the model which implies four parameters to be determined.

The gradients were first computed with the three outliers but we found that they had too much weight and therefore we show below the gradients computed after removing the three outliers.

We find a vertical metallicity gradient of $-$0.415~$\pm$~0.030~dex kpc$^{-1}$, steeper but in agreement, within the uncertainties, with another value derived from the OCs, $-$0.385~$\pm$~0.026~dex kpc$^{-1}$ \citep{Joshi+2024}. Other determinations based on the OCs give shallower gradients: $-$0.340~$\pm$~0.030~dex kpc$^{-1}$ \citep{Piatti+1995}, $-$0.295~$\pm$~0.050~dex kpc$^{-1}$ \citep{Chen+2003}, $-$0.252~$\pm$~0.039~dex kpc$^{-1}$ \citep{Zhong+2020}, $-$0.031~$\pm$~0.077~dex kpc$^{-1}$ \citep{Yang+2025}. Other tracers tend to give slightly shallower gradients, for instance $-$0.30~$\pm$~0.03~dex kpc$^{-1}$  from field giant stars \citep{Soubiran+2008} and $-$0.243~$_{-0.053}^{+0.039}$~dex kpc$^{-1}$ from G dwarfs \citep{Schlesinger+2014}. The metallicity dispersion is about 0.17 dex at |Z|=0 and decreases to less than 0.1 dex at |Z|=1 kpc.

For the radial direction, the fit confirms the two slopes with a nearly flat gradient in the outer disc, but the exact position of the change in slope is not well defined. The most probable position of the knee is at R$_{\text{G}}$=12.43$\pm$0.04 kpc, but there is another solution at R$_{\text{G}}$=10.80$\pm$0.04 kpc. The linear fit on the inner radial distribution gives a metallicity gradient of $-$0.084~$\pm$~0.004~dex kpc$^{-1}$, shown in the left panel of Fig~\ref{fig:Z_MFeh_relationships}. This value is on the steeper side compared to many previous determinations based on OCs which range between $-$0.048 and $-$0.077~dex kpc$^{-1}$ \citep{Netopil+2016, Chen+2003, Carrera+2019, Donor+2020, Spina+2021b, Zhang+2021, Spina+2022, Yang+2025}. Our value is in good agreement with the recent determination by \cite{Haywood+2024} who found a value of $-$0.080~$\pm$~0.003~dex kpc$^{-1}$ between 6 and 11 kpc based on a sample of red giants in the thin disc from APOGEE. They found that the gradient flattens at R$_{\text{G}} >$ 11-12 kpc, also in good agreement with our sample.

We investigate in the next section whether the radial gradient varies when the sample is divided into age bins.

\subsection{Age dependence of the radial metallicity gradient}

\begin{figure}[htbp]
	\centering
    \includegraphics[angle=0,width=90mm]{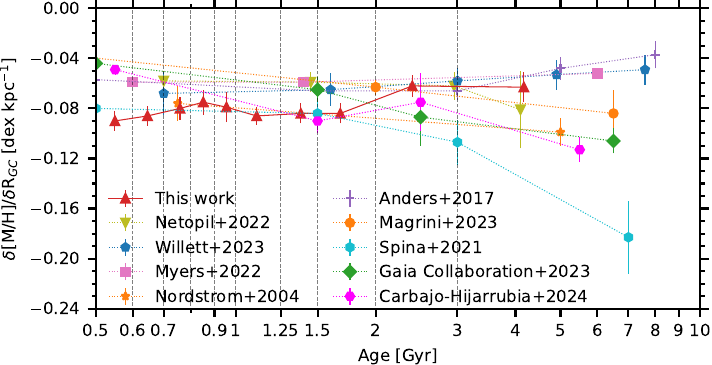}
	\caption{Evolution of the radial metallicity gradients with age (data from Table.~\ref{tab:age_gradients}) and comparison with the literature. The median age of the OCs in each age bin is taken as a horizontal coordinate. }
	\label{fig:metallicity_gradients_vs_age}
\end{figure}

In order to investigate the evolution of the radial metallicity gradient, we divided our OC sample into ten age bins, and we computed the corresponding gradient in R$_{\text{G}}$ 
in each bin with the same linear fitting method as previously, with a Bayesian linear gradient + scatter model. Due to the flattening of the gradient in the outer disc, the fit was performed on clusters with R$_{\text{G}}$ < 12.5 kpc. The age bins were chosen to ensure a similar number of OCs in each. The resulting slopes and dispersions are listed in Table~\ref{tab:age_gradients}. The values of the gradient are also shown in  Fig.~\ref{fig:metallicity_gradients_vs_age} together with other studies for comparison. The radial gradient starts with a steep value in the two youngest age bins, about $-$0.090~dex kpc$^{-1}$, sightly lower than all other values from the literature. From 700 Myr to 2 Gyr, the gradient remains at around $-$0.08 dex kpc$^{-1}$ in better agreement with the other studies. It flattens at about $-$0.06 dex kpc$^{-1}$ in the two oldest bins in good agreement with \citet{Anders+2017}, \citet{Myers+2022}, and \citet{Willett+2023}, but with an increasing dispersion that was below 0.1 dex in the younger bins.

Figure ~\ref{fig:metallicity_gradients_vs_age} shows a wide range of literature values at ages older than 3 Gyr, from very steep gradients determined by \cite{Spina+2021b} with OCs, to flatter gradients found by us and some others. In agreement with us, \citet{Willett+2023} and \cite{Anders+2017} found a flattening of the radial metallicity gradient leveraging oscillating red giants known to provide reliable ages while \citep{Myers+2022} also used OCs. \cite{Nordstr+2004} found steeper trends from F and G dwarfs, \cite{Spina+2021b, Netopil+2022, Magrini+2023, Gaia+2023, Carbajo-Hijarrubia+2024} from OC. This lack of consensus among the literature results at the oldest 
ages likely reflects uncertainties of ages and metallicities depending on the tracers used, as well as differences in the boundaries of the samples in age and distance, selection biases, and insufficient statistics.

\begin{table}[htbp]
\centering
\caption{Radial metallicity gradient for OCs with R$_{\text{G}}$~$\leq$~12.5~kpc in ten age bins.}
\label{tab:age_gradients}
\begin{tabular}{lccc}
\toprule
Age  &  $\frac{d[\mathrm{M/H}]}{dR_{\text{G}}}$ &  $\sigma$ & N \\
(Gyr) & (dex kpc$^{-1}$) & (dex)& \\
\midrule
0.5-0.6 &  $-0.090 \pm 0.008$ &  $0.088 \pm 0.011$  & 53  \\
0.6-0.7 &  $-0.086 \pm 0.008$ &  $0.085 \pm 0.010$ & 61  \\
0.7-0.8 &  $-0.080 \pm 0.009$ &  $0.089 \pm 0.011$ & 55  \\
0.8-0.9 &  $-0.075 \pm 0.010$ &  $0.090 \pm 0.013$ & 45 \\
0.9-1.0 &   $-0.079 \pm 0.012$ &  $0.096 \pm 0.014$ &  39  \\
1.0-1.25 &   $-0.086 \pm 0.007$ &  $0.094 \pm 0.010$ &  80  \\
1.25-1.5 &   $-0.084 \pm 0.008$ &  $0.085 \pm 0.011$ &  61  \\
1.5-2.0 &  $-0.084 \pm 0.008$ &   $0.097 \pm 0.011$ &  65  \\
2.0-3.0 &   $-0.062 \pm 0.009$ &  $0.106 \pm 0.013$ &  52  \\
3.0-10.0 &   $-0.063 \pm 0.012$ &  $0.140 \pm 0.017$ &  41  \\
\bottomrule
\end{tabular}
\tablefoot{$\sigma$ is the intrinsic [M/H] dispersion and N is the number of fitted clusters in each age bin.}
\end{table}

\subsection{Age-metallicity relation}

The AMR is a fundamental observational constraint to any theory of chemical evolution of the Galactic disc. No evidence of a variation of metallicity with age has been found from OCs by \cite{ Netopil+2016} while a very weak trend is reported by \cite{Pancino+2010, Zhong+2020,Joshi+2024}. The AMR is characterised by a large scatter at all ages, whatever the tracer, and samples of old OCs have always been too small to provide conclusive results. Now with our unprecedentedly large and homogeneous sample, we can re-investigate the AMR in the light of the R$_{\text{G}}$ dependence reported in the previous section.

\begin{figure}[htbp]
	\centering
 \includegraphics[angle=0,width=88mm]{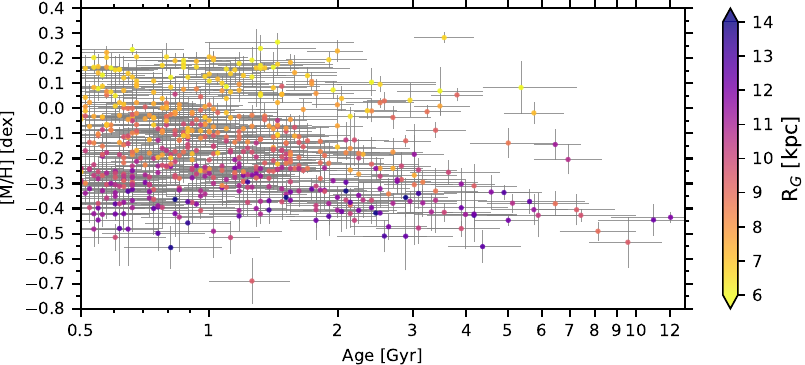} 	
 \caption{Metallicity vs age distribution of the 597 OCs colour-coded by their galactocentric radius.}
	\label{fig:AMR_1}
\end{figure}

We plot the metallicity versus age  distribution of the sample in Fig. \ref{fig:AMR_1} colour-coded by R$_{\text{G}}$. The scatter in metallicity is clearly visible at all ages, but the colouring shows that the metallicities are not randomly distributed, but ordered according to R$_{\text{G}}$. This is another way to visualise the radial gradient. The distribution is well populated up to $\sim$1.5 Gyr, with  metallicities spanning the full range from $-$0.55 to +0.25. As noted previously, along the age range the metal-rich clusters dominate at low R$_{\text{G}}$ (in yellow), while the decreasing metallicity corresponds to increasing R$_{\text{G}}$ (from yellow to purple). For ages older than 1.5 Gyr, there are two sequences. Metal-poor OCs are more represented and define a clear sequence delimited at the bottom by a metallicity limit of [M/H]$\simeq$$-$0.55, constant along the age axis. This sequence also has a sharp edge above which there are fewer clusters. Above that edge, more metal-rich OCs form a distinct sequence, less populated and more dispersed in metallicity. It is likely that we face a survival bias at the oldest ages because a large part of the OCs have been destroyed, in particular those born in an inner radius.

\begin{figure}[htbp]
	\centering
 \includegraphics[angle=0,width=66mm]{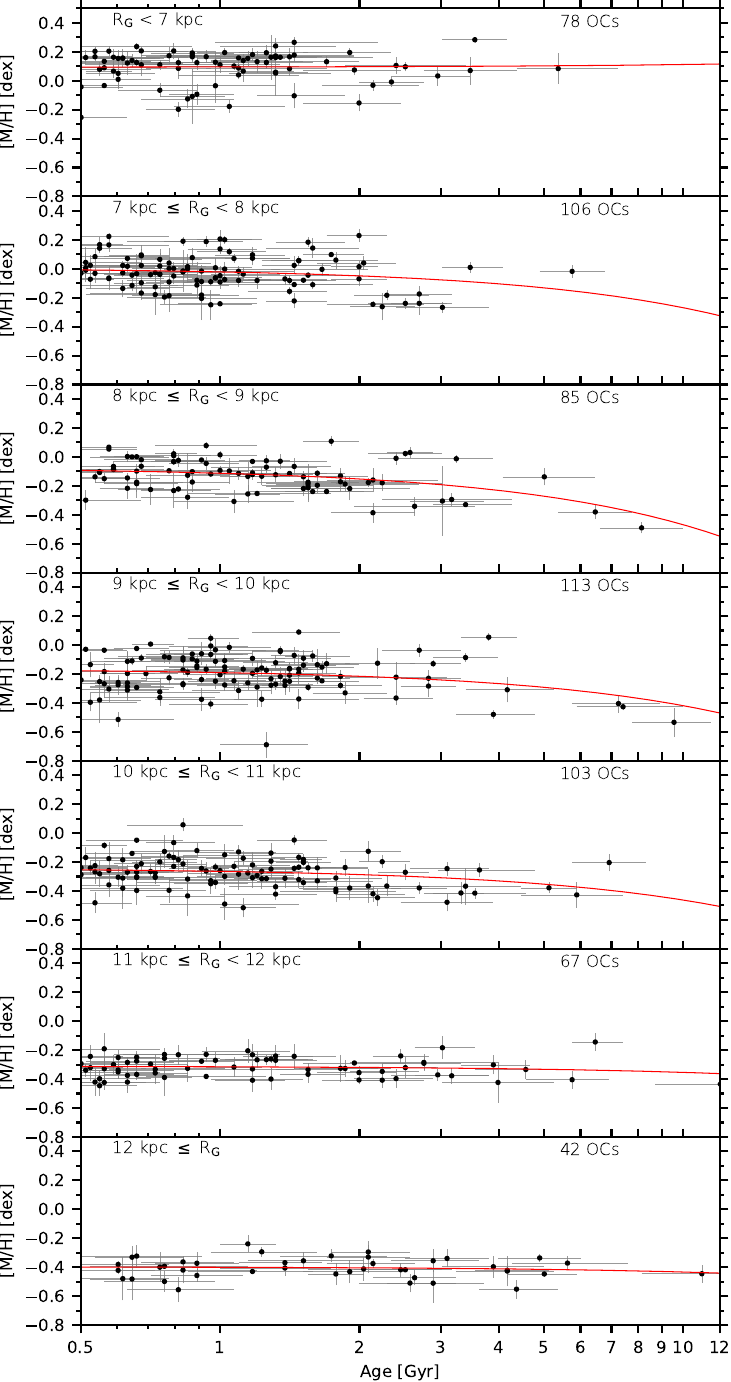}
	\caption{AMR traced by old OCs at different distances from the Galactic centre. The linear fit in red highlights the trend (metallicity and age uncertainties have been taken into account via MC sampling). The three outliers mentioned earlier have not been considered.}
	\label{fig:AMR_2}
\end{figure}

As the final representation of our sample, we now plot in Fig. \ref{fig:AMR_2} the age-metallicity distribution in seven bins of R$_{\text{G}}$, and fit a line representing the AMR in each bin. The summary of the AMR in the seven R$_{\text{G}}$ bins is given in Table \ref{tab:dispersion} in the form of a slope, a scatter (MAD) and the mean [M/H] at 500 Myr. We see a clear offset of the mean metallicity from one bin to another. Starting from the smallest radius bin, centred around [M/H]=+0.09, the mean metallicity is shifted downward  in each radius bin by an amount ranging from 0.02 dex to nearly 0.1 dex, down to [M/H]=-0.40 in the most distant bin. The value of the slope tells us whether there was a chemical enrichment at a given radius during the considered time range. For the bins between  7 and 11 kpc, the value of the slope is slightly negative, indicating that the metallicity increases with time. In the smallest radius bin, corresponding to the inner disc at R$_{\text{G}}<$7 kpc, and in the two last bins, corresponding to the outer disc at R$_{\text{G}}\geq$11 kpc, the slope is compatible with zero, indicating a metallicity nearly constant from 0.5 to 10 Gyr. The dispersion in each bin reflects both the homogeneity of the matter from which the clusters formed, and the strength of the radial mixing. The lowest dispersion is observed in the first bin and the two last bins, where the metallicity was also found to be constant, reflecting to the homogeneity and stability of the medium. In between, from 7 to 11 kpc, the larger dispersion and increasing metallicity with time indicate that star formation and chemical enrichment took place.  The change of regime that we see at 7 kpc and 11 kpc is possibly related to the dynamical effect of the bar described by \cite{Haywood+2024}.

The main outcome of our study is that the large dispersion of the AMR, when all the OCs are considered together, results from the superposition of populations at different radii, each with its own AMR. Our results are nicely consistent with the model of \cite{Minchev+2018}, and predict that the AMR is shifted downward for outer radii, which is the result of the negative radial metallicity gradient of the interstellar matter that was present in the disc when the clusters formed. The bottom metallicity of [M/H]$\simeq$$-$0.55 that we observe at all ages is relatively high, indicating that most of the chemical enrichment of the Galactic disc occurred  before the OCs were formed.

\begin{table}[htbp]
\centering
\caption{AMR in different bins of galactocentric distance. }
\label{tab:dispersion}
\begin{tabular}{cccc}
\toprule
R$_{\text{G}}$ &  $\frac{d[\mathrm{M/H}]}{d{\mathrm Age}}$  & MAD &  [M/H]$_{\rm 0.5Gyr}$  \\
kpc & dex Gyr$^{-1}$ & dex & dex  \\
\midrule
< 7 & $+0.002 \pm 0.016$ & 0.050  &   $+0.094$ \\
7 - 8 & $-0.028 \pm 0.016$ & 0.066 &   $-0.008$\\
8 - 9 & $-0.040 \pm 0.010$ & 0.074 &   $-0.093$\\
9 - 10 & $-0.025 \pm 0.010$ & 0.080  &   $-0.178$ \\
10 - 11 & $-0.022 \pm 0.010$ & 0.058 &  $-0.254$ \\
11 - 12 & $-0.004 \pm 0.006$ & 0.054 &  $-0.313$\\
$\geq$ 12 & $-0.004\pm 0.008$ & 0.047 &  $-0.399$ \\
\bottomrule
\end{tabular}
\tablefoot{The slope of the linear fitting, represented in Fig~\ref{fig:AMR_2}, with the dispersion and the intercept metallicity at 0.5 Gyr are provided.}
\end{table}

\section{Summary and conclusion}
In this work we demonstrated that metallicities leveraging Gaia XP spectra combined with infrared photometry are of high enough quality to study old OCs. We found the best metallicities to be those deduced from the vetted sample of bright red giants of \cite{Andrae+2023b}. They give OC metallicities in good agreement with those deduced from high-resolution spectroscopy. Based on this catalogue of stellar parameters, we determined metallicities for the largest ever sample of OCs older than 500 Myr, containing $\sim$600 clusters. This is also the first all-sky catalogue of OC metallicities with a typical precision of 0.05 dex. Most clusters were poorly studied before, and some were recently discovered by \cite{Hunt+2023}. Several OCs were found to have properties making them appear as outliers compared to the bulk of the sample and they deserve further investigation. Only a very few clusters were found to have metallicities below [M/H]=$-$0.55 or above [M/H]=+0.25. The bulk of the sample extends from 5.5 to 15 kpc in galactentric radius, and up to 1.2 kpc perpendicularly to the Galactic mid-plane. We are aware that our sample of OCs is affected by complex selection effects and that the population of OCs we see today is strongly affected by a survival bias since many clusters have been destroyed, in particular those born in the inner disc where the gravitational influence of the Milky Way is stronger. It is also likely that many clusters have not yet been detected. 
Despite this incompleteness, the maps clearly show the dependence of the metallicity distribution of old OCs with age and Galactic position. We measured the radial and vertical gradients in agreement with previous determinations in the literature. In line with previous studies, we find two regimes in the radial gradient with a change in slope at around R$_{\text{G}}$=10-12 kpc, being nearly flat in the outer disc. The inner gradient evolves with time, being steep in the youngest age bins and shallower for older clusters. Metal-rich OCs dominate in the inner disc, while metal-poor clusters dominate outside the solar circle, whatever their age, and without significant radial mixing. 

The larger scale height of the outer clusters, which are also more metal-poor and older on average, is consistent with the gradual thinning of the disc or the heating of the oldest OCs by dynamical processes. Finally, we investigated the AMR of the OCs and showed that the large dispersion at all ages is related to the clusters' radii. When the sample is divided into bins of galactocentric distance, one can see a less scattered AMR in each bin, with the mean metallicity decreasing regularly with increasing radius. Beyond 11 kpc and closer than 7 kpc from the Galactic centre, the metallicity distribution is remarkably homogeneous at all ages. Our results show that most of the chemical enrichment of the disc had already taken place when the oldest surviving clusters formed about 10 Gyr ago. Our findings agree with the metallicity distribution predicted by the model of \citet{Minchev+2018} in which the AMR is shifted downward for outer radii as the result of the negative radial metallicity gradient of the interstellar matter.

\section{Data availability}

Table~\ref{table:metallicity} is only available in electronic form at the CDS via anonymous ftp to cdsarc.u-strasbg.fr (130.79.128.5) or via http://cdsweb.u-strasbg.fr/cgi-bin/qcat?J/A+A/.

\begin{acknowledgements}

The authors thank an anonymous reviewer for valuable suggestions that improved our results and presentation. The authors also would like to thank Laia Casamiquela and Misha Haywood for valuable discussions. This work is supported by the National Natural Science Foundation of China (NSFC) under grant 12303037, the Natural Science Foundation of Sichuan Province (No. 2025ZNSFSC0879), and the Fundamental Research Funds of China West Normal University (CWNU, No.23KE024). Qingshun Hu would like to acknowledge the financial support provided by the China Scholarship Council program (Grant No. 202308510136). This study has made an indirect use of Gaia data, through large catalogues of open clusters and stellar parameters leveraring astrometric, photometric and spectrophotometric data published in the third data release. Gaia is operated by the European Space Agency (ESA) (\url{https://www.cosmos.esa.int/gaia}). The preparation of this work has made extensive use of Topcat \citep{Taylor+2005}, of the Simbad and VizieR databases at CDS, Strasbourg, France, and of NASA's Astrophysics Data System Bibliographic Services.

\end{acknowledgements}



\begin{thebibliography}{}
	
	
	\bibitem[Abdurro'uf et al.(2022)]{Abdurro'uf+2022} Abdurro'uf, Accetta, K., Aerts, C., et al.\ 2022, \apjs, 259, 35
	
	\bibitem[Anders et al.(2017)]{Anders+2017} Anders, F., Chiappini, C., Minchev, I., et al.\ 2017, \aap, 600, A70
	
	\bibitem[Anders et al.(2022)]{Anders+2022} Anders, F., Khalatyan, A., Queiroz, A.~B.~A., et al.\ 2022, \aap, 658, A91
	
	\bibitem[Andrae et al.(2023)]{Andrae+2023a} Andrae, R., Fouesneau, M., Sordo, R., et al.\ 2023, \aap, 674, A27
	
	\bibitem[Andrae et al.(2023)]{Andrae+2023b} Andrae, R., Rix, H.-W., \& Chandra, V.\ 2023, \apjs, 267, 8
	
	\bibitem[Buder et al.(2021)]{Buder+2021} Buder, S., Sharma, S., Kos, J., et al.\ 2021, \mnras, 506, 150
	
	\bibitem[Cantat-Gaudin et al.(2020)]{Cantat-Gaudin+2020} Cantat-Gaudin, T., Anders, F., Castro-Ginard, A., et al.\ 2020, \aap, 640, A1
	
	\bibitem[Cantat-Gaudin \& Casamiquela(2024)]{Cantat-Gaudin+2024} Cantat-Gaudin, T. \& Casamiquela, L.\ 2024, \nar, 99, 101696
	
	\bibitem[Carbajo-Hijarrubia et al.(2024)]{Carbajo-Hijarrubia+2024} Carbajo-Hijarrubia, J., Casamiquela, L., Carrera, R., et al.\ 2024, \aap, 687, A239
	
	\bibitem[Carrera et al.(2019)]{Carrera+2019} Carrera, R., Bragaglia, A., Cantat-Gaudin, T., et al.\ 2019, \aap, 623, A80
	
	\bibitem[Casamiquela et al.(2016)]{Casamiquela+2016} Casamiquela, L., Carrera, R., Jordi, C., et al.\ 2016, \mnras, 458, 3150
	
	\bibitem[Casamiquela et al.(2021)]{Casamiquela+2021} Casamiquela, L., Soubiran, C., Jofr{\'e}, P., et al.\ 2021, \aap, 652, A25
	
	\bibitem[Cavallo et al.(2024)]{Cavallo+2024} Cavallo, L., Spina, L., Carraro, G., et al.\ 2024, \aj, 167, 12
	
	\bibitem[Chambers et al.(2016)]{Chambers+2016} Chambers, K.~C., Magnier, E.~A., Metcalfe, N., et al.\ 2016, arXiv:1612.05560
	
	\bibitem[Chen et al.(2003)]{Chen+2003} Chen, L., Hou, J.~L., \& Wang, J.~J.\ 2003, \aj, 125, 1397
	
	\bibitem[Cirasuolo et al.(2014)]{Cirasuolo+2014} Cirasuolo, M., Afonso, J., Carollo, M., et al.\ 2014, \procspie, 9147, 91470N
	
	\bibitem[Cropper et al.(2018)]{Cropper+2018} Cropper, M., Katz, D., Sartoretti, P., et al.\ 2018, \aap, 616, A5
	
	\bibitem[Cutri et al.(2003)]{Cutri+2003} Cutri, R.~M., Skrutskie, M.~F., van Dyk, S., et al.\ 2003, 2MASS All Sky Catalog of point sources.
	
	\bibitem[Cutri et al.(2013)]{Cutri+2013} Cutri, R.~M., Wright, E.~L., Conrow, T., et al.\ 2013, Explanatory Supplement to the AllWISE Data Release Products, by R. M. Cutri et al., Explanatory Supplement to the AllWISE Data Release Products, 1
	
	 \bibitem[De Angeli et al.(2023)]{DeAngeli+2023} De Angeli, F., Weiler, M., Montegriffo, P., et al.\ 2023, \aap, 674, A2
	 
	 \bibitem[de Jong et al.(2019)]{deJong+2019} de Jong, R.~S., Agertz, O., Berbel, A.~A., et al.\ 2019, 175, 3
	 
	 \bibitem[De Silva et al.(2015)]{DeSilva+2015} De Silva, G.~M., Freeman, K.~C., Bland-Hawthorn, J., et al.\ 2015, \mnras, 449, 2604
	 
	 \bibitem[Dias et al.(2021)]{Dias+2021} Dias, W.~S., Monteiro, H., Moitinho, A., et al.\ 2021, \mnras, 504, 356
	 
	 \bibitem[Donor et al.(2020)]{Donor+2020} Donor, J., Frinchaboy, P.~M., Cunha, K., et al.\ 2020, \aj, 159, 199
	 
	 \bibitem[Eilers et al.(2022)]{Eilers+2022} Eilers, A.-C., Hogg, D.~W., Rix, H.-W., et al.\ 2022, \apj, 928, 23
	 
	 \bibitem[Fallows \& Sanders(2024)]{Fallows+2024} Fallows, C.~P. \& Sanders, J.~L.\ 2024, \mnras, 531, 2126
	 
	 \bibitem[Friel(1995)]{Friel+1995} Friel, E.~D.\ 1995, \araa, 33, 381
	 
	 \bibitem[Friel et al.(2002)]{Friel+2002} Friel, E.~D., Janes, K.~A., Tavarez, M., et al.\ 2002, \aj, 124, 2693
	 
	 \bibitem[Fu et al.(2022)]{Fu+2022} Fu, X., Bragaglia, A., Liu, C., et al.\ 2022, \aap, 668, A4
	 
	 \bibitem[Gaia Collaboration et al.(2021)]{Gaia+2021} Gaia Collaboration, Brown, A.~G.~A., Vallenari, A., et al.\ 2021, \aap, 649, A1
	 
	 \bibitem[Gaia Collaboration et al.(2016)]{Gaia+2016} Gaia Collaboration, Prusti, T., de Bruijne, J.~H.~J., et al.\ 2016, \aap, 595, A1
	 
	 \bibitem[Gaia Collaboration et al.(2023)]{GaiaCollRecio-Blanco+2023} Gaia Collaboration, Recio-Blanco, A., Kordopatis, G., et al.\ 2023, \aap, 674, A38
	 
	 \bibitem[Gaia Collaboration et al.(2023)]{Gaia+2023} Gaia Collaboration, Vallenari, A., Brown, A.~G.~A., et al.\ 2023, \aap, 674, A1
	 
	 \bibitem[Garro et al.(2023)]{Garro+2023} Garro, E.~R., Fern{\'a}ndez-Trincado, J.~G., Minniti, D., et al.\ 2023, \aap, 669, A136
	 
	 \bibitem[GRAVITY Collaboration et al.(2018)]{GRAVITY+2018} GRAVITY Collaboration, Abuter, R., Amorim, A., et al.\ 2018, \aap, 615, L15
	 
	 \bibitem[Haywood et al.(2024)]{Haywood+2024} Haywood, M., Khoperskov, S., Cerqui, V., et al.\ 2024, \aap, 690, A147
	 
	 \bibitem[Heiter et al.(2014)]{Heiter+2014} Heiter, U., Soubiran, C., Netopil, M., et al.\ 2014, \aap, 561, A93
	 
	 \bibitem[Hunt \& Reffert(2023)]{Hunt+2023} Hunt, E.~L. \& Reffert, S.\ 2023, \aap, 673, A114
	 
	 \bibitem[Hunt \& Reffert(2024)]{Hunt+2024} Hunt, E.~L. \& Reffert, S.\ 2024, \aap, 686, A42
	 
	 \bibitem[Janes(1979)]{Janes+1979} Janes, K.~A.\ 1979, \apjs, 39, 135
	 
	 \bibitem[Jin et al.(2024)]{Jin+2024} Jin, S., Trager, S.~C., Dalton, G.~B., et al.\ 2024, \mnras, 530, 2688
	 
	 \bibitem[Joshi et al.(2024)]{Joshi+2024} Joshi, Y.~C., Deepak, \& Malhotra, S.\ 2024, Frontiers in Astronomy and Space Sciences, 11, 1348321
	 
	 \bibitem[Katz et al.(2023)]{Katz+2023} Katz, D., Sartoretti, P., Guerrier, A., et al.\ 2023, \aap, 674, A5
	 
	 \bibitem[Khalatyan et al.(2024)]{Khalatyan+2024} Khalatyan, A., Anders, F., Chiappini, C., et al.\ 2024, \aap, 691, A98
	 
	 \bibitem[Kos et al.(2025)]{Kos+2025} Kos, J., Buder, S., Beeson, K.~L., et al.\ 2025, arXiv:2501.06140
	 
	 \bibitem[Magrini et al.(2023)]{Magrini+2023} Magrini, L., Viscasillas V{\'a}zquez, C., Spina, L., et al.\ 2023, \aap, 669, A119
	 
	 \bibitem[Marocco et al.(2021)]{Marocco+2021} Marocco, F., Eisenhardt, P.~R.~M., Fowler, J.~W., et al.\ 2021, \apjs, 253, 8
	 
	 \bibitem[Minchev et al.(2018)]{Minchev+2018} Minchev, I., Anders, F., Recio-Blanco, A., et al.\ 2018, \mnras, 481, 1645
	 
	 \bibitem[Montegriffo et al.(2023)]{Montegriffo+2023} Montegriffo, P., De Angeli, F., Andrae, R., et al.\ 2023, \aap, 674, A3
	 
	 
	 \bibitem[Myers et al.(2022)]{Myers+2022} Myers, N., Donor, J., Spoo, T., et al.\ 2022, \aj, 164, 85
	 
	 \bibitem[Netopil et al.(2022)]{Netopil+2022} Netopil, M., Oralhan, {\.I}. A., {\c{C}}akmak, H., et al.\ 2022, \mnras, 509, 421
	 
	 \bibitem[Netopil et al.(2016)]{Netopil+2016} Netopil, M., Paunzen, E., Heiter, U., et al.\ 2016, \aap, 585, A150
	 
	 \bibitem[Nordstr{\"o}m et al.(2004)]{Nordstr+2004} Nordstr{\"o}m, B., Mayor, M., Andersen, J., et al.\ 2004, \aap, 418, 989
	 
	 \bibitem[Onken et al.(2019)]{Onken+2019} Onken, C.~A., Wolf, C., Bessell, M.~S., et al.\ 2019, \pasa, 36, e033
	 
	 \bibitem[Palla et al.(2024)]{Palla+2024} Palla, M., Magrini, L., Spitoni, E., et al.\ 2024, \aap, 690, A334
	 
	 \bibitem[Pancino et al.(2010)]{Pancino+2010} Pancino, E., Carrera, R., Rossetti, E., et al.\ 2010, \aap, 511, A56
	 
	 \bibitem[Piatti et al.(1995)]{Piatti+1995} Piatti, A.~E., Claria, J.~J., \& Abadi, M.~G.\ 1995, \aj, 110, 2813
	 
	 \bibitem[Queiroz et al.(2018)]{Queiroz+2018} Queiroz, A.~B.~A., Anders, F., Santiago, B.~X., et al.\ 2018, \mnras, 476, 2556
	 
	 \bibitem[Randich et al.(2022)]{Randich+2022} Randich, S., Gilmore, G., Magrini, L., et al.\ 2022, \aap, 666, A121
	 
	 
	 \bibitem[Recio-Blanco et al.(2023)]{Recio-Blanco+2023} Recio-Blanco, A., de Laverny, P., Palicio, P.~A., et al.\ 2023, \aap, 674, A29
	
	\bibitem[Renaud et al.(2025)]{Renaud+2025} Renaud, F., Ratcliffe, B., Minchev, I., et al.\ 2025, \aap, 694, A56
	
	\bibitem[Schlesinger et al.(2014)]{Schlesinger+2014} Schlesinger, K.~J., Johnson, J.~A., Rockosi, C.~M., et al.\ 2014, \apj, 791, 112
	
	\bibitem[Soubiran et al.(2008)]{Soubiran+2008} Soubiran, C., Bienaym{\'e}, O., Mishenina, T.~V., et al.\ 2008, \aap, 480, 91
	
	\bibitem[Spina et al.(2022)]{Spina+2022} Spina, L., Magrini, L., \& Cunha, K.\ 2022, Universe, 8, 87
	
	
	\bibitem[Spina et al.(2021)]{Spina+2021b} Spina, L., Ting, Y.-S., De Silva, G.~M., et al.\ 2021, \mnras, 503, 3279
	
	
	\bibitem[Steinmetz et al.(2020)]{Steinmetz+2020} Steinmetz, M., Matijevi{\v{c}}, G., Enke, H., et al.\ 2020, \aj, 160, 82
	
	\bibitem[Steinmetz et al.(2006)]{Steinmetz+2006} Steinmetz, M., Zwitter, T., Siebert, A., et al.\ 2006, \aj, 132, 1645
	
	
	\bibitem[Tarricq et al.(2021)]{Tarricq+2021} Tarricq, Y., Soubiran, C., Casamiquela, L., et al.\ 2021, \aap, 647, A19
	
	\bibitem[Taylor(2005)]{Taylor+2005} Taylor, M.~B.\ 2005, Astronomical Data Analysis Software and Systems XIV, 347, 29
	
	
	\bibitem[Vickers et al.(2021)]{Vickers+2021} Vickers, J.~J., Shen, J., \& Li, Z.-Y.\ 2021, \apj, 922, 189
	
	\bibitem[Wang et al.(2022)]{Wang+2022} Wang, C., Huang, Y., Yuan, H., et al.\ 2022, \apjs, 259, 51
	
	\bibitem[Willett et al.(2023)]{Willett+2023} Willett, E., Miglio, A., Mackereth, J.~T., et al.\ 2023, \mnras, 526, 2141
	
	\bibitem[Yang et al.(2025)]{Yang+2025} Yang, G., Zhao, J., Yang, Y., et al.\ 2025, \aj, 169, 214
	
	\bibitem[Yanny et al.(2009)]{Yanny+2009} Yanny, B., Rockosi, C., Newberg, H.~J., et al.\ 2009, \aj, 137, 4377
	
	\bibitem[Ye et al.(2025)]{Ye+2025} Ye, X., Wu, W., Allende Prieto, C., et al.\ 2025, \aap, 695, A75
	
	\bibitem[Zhang et al.(2021)]{Zhang+2021} Zhang, R., Lucatello, S., Bragaglia, A., et al.\ 2021, \aap, 654, A77
	
	\bibitem[Zhang et al.(2024)]{Zhang+2024} Zhang, R., Wang, G.-J., Lu, Y., et al.\ 2024, \aap, 692, A212
	
	\bibitem[Zhang et al.(2023)]{Zhang+2023} Zhang, X., Green, G.~M., \& Rix, H.-W.\ 2023, \mnras, 524, 1855
	
	\bibitem[Zhong et al.(2020)]{Zhong+2020} Zhong, J., Chen, L., Wu, D., et al.\ 2020, \aap, 640, A127
	
	
	
\end{thebibliography}


\begin{appendix}

\FloatBarrier
\onecolumn

\begin{sidewaystable}[h!]\tiny 
\section{Metallicity parameters derived from six catalogues}
\centering
\begin{threeparttable}
	\caption{Metallicity parameters per cluster derived from the six catalogues.}\label{table:metallicity}
	\begin{tabular}{c c c c c c c c c c c c c c c c c c c c}
		\hline\noalign{\smallskip}
		\hline\noalign{\smallskip}
 \multirow{3}{*}{} & \multicolumn{3}
 {c|}{\citet{Anders+2022}} &  \multicolumn{3}
 {c|}{\citet[][Full]{Andrae+2023b}} &   \multicolumn{3}
 {c|}{\citet[][Giants]{Andrae+2023b}} &   \multicolumn{3}
 {c|}{\citet{Zhang+2023}} &  \multicolumn{3}
 {c|}{\citet{Khalatyan+2024}}   &  \multicolumn{3}
{c}{\citet{Fallows+2024}}  \\
		\hline\noalign{\smallskip}
        
	Cluster &   [M/H] & $\sigma_{\text{[M/H]}}$ & N & [M/H] & $\sigma_{\text{[M/H]}}$ & N &  [M/H] & $\sigma_{\text{[M/H]}}$  & N &   [Fe/H] & $\sigma_{\text{[Fe/H]}}$ & N &   [M/H] & $\sigma_{\text{[M/H]}}$ & N & [Fe/H] & $\sigma_{\text{[Fe/H]}}$ & N \\
		\hline\noalign{\smallskip}

ASCC 61 & $-$0.28 & 0.16 & 43 & $-$0.22 & 0.12 & 72 & $-$0.08 & 0.03 & 14 & $-$0.25 & 0.10 & 50 & $-$0.25 & 0.08 & 81 & $-$0.14 & 0.08 & 71 \\
ASCC 63 & $-$0.04 & 0.14 & 7 & $-$0.44 & 0.18 & 34 & $-$0.14 & 0.06 & 8 & -- & -- & -- & $-$0.13 & 0.07 & 18 & $-$0.35 & 0.23 & 29 \\
ASCC 75 & $-$0.16 & 0.07 & 7 & $-$0.21 & 0.10 & 6 & -- & -- & -- & $-$0.30 & 0.14 & 10 & $-$0.18 & 0.03 & 13 & -- & -- & -- \\
Alessi 22 & $-$0.16 & 0.08 & 16 & $-$0.08 & 0.08 & 22 & -- & -- & -- & 0.07 & 0.07 & 13 & $-$0.02 & 0.07 & 17 & $-$0.06 & 0.09 & 29 \\
Alessi 50 & $-$0.18 & 0.09 & 90 & $-$0.51 & 0.14 & 54 & $-$0.38 & 0.04 & 5 & $-$0.41 & 0.07 & 16 & $-$0.37 & 0.09 & 25 & $-$0.34 & 0.18 & 51 \\
Alessi 62 & $-$0.07 & 0.09 & 126 & 0.01 & 0.07 & 165 & -- & -- & -- & 0.05 & 0.07 & 114 & $-$0.05 & 0.09 & 167 & $-$0.00 & 0.04 & 199 \\
Alessi 116 & $-$0.20 & 0.09 & 98 & $-$0.11 & 0.06 & 129 & $-$0.06 & 0.02 & 7 & $-$0.07 & 0.08 & 88 & $-$0.10 & 0.08 & 112 & $-$0.04 & 0.06 & 124 \\
Alessi 170 & $-$0.14 & 0.10 & 128 & $-$0.24 & 0.14 & 111 & 0.01 & 0.02 & 6 & $-$0.05 & 0.11 & 70 & $-$0.08 & 0.06 & 80 & $-$0.07 & 0.07 & 97 \\
Auner 1 & $-$0.38 & 0.12 & 45 & $-$0.63 & 0.13 & 21 & $-$0.42 & 0.05 & 5 & $-$0.66 & 0.12 & 13 & $-$0.46 & 0.09 & 28 & $-$0.51 & 0.16 & 26 \\
BH 55 & $-$0.08 & 0.12 & 106 & $-$0.37 & 0.12 & 73 & $-$0.27 & 0.04 & 11 & $-$0.36 & 0.11 & 13 & $-$0.29 & 0.07 & 44 & $-$0.23 & 0.11 & 72 \\
BH 66 & $-$0.27 & 0.13 & 54 & $-$0.50 & 0.18 & 94 & $-$0.28 & 0.08 & 17 & $-$0.50 & 0.09 & 20 & $-$0.33 & 0.08 & 48 & $-$0.32 & 0.16 & 77 \\
BH 73 & $-$0.14 & 0.15 & 26 & $-$0.39 & 0.08 & 24 & -- & -- & -- & $-$0.44 & 0.10 & 10 & $-$0.32 & 0.06 & 24 & $-$0.34 & 0.14 & 28 \\
BH 84 & $-$0.38 & 0.18 & 11 & $-$0.40 & 0.17 & 30 & -- & -- & -- & -- & -- & -- & $-$0.18 & 0.07 & 9 & $-$0.50 & 0.14 & 28 \\
BH 85 & $-$0.16 & 0.08 & 112 & $-$0.47 & 0.18 & 38 & $-$0.29 & 0.04 & 8 & $-$0.34 & 0.08 & 18 & $-$0.42 & 0.13 & 56 & $-$0.27 & 0.12 & 40 \\
BH 118 & $-$0.19 & 0.11 & 68 & $-$0.29 & 0.15 & 42 & $-$0.22 & 0.08 & 14 & $-$0.41 & 0.15 & 28 & $-$0.20 & 0.11 & 49 & $-$0.21 & 0.10 & 42 \\
BH 202 & $-$0.27 & 0.11 & 153 & $-$0.19 & 0.10 & 124 & 0.16 & 0.03 & 10 & $-$0.36 & 0.10 & 127 & $-$0.12 & 0.06 & 252 & $-$0.27 & 0.10 & 134 \\
BH 211 & $-$0.22 & 0.12 & 64 & $-$0.09 & 0.20 & 13 & 0.21 & 0.02 & 5 & $-$0.40 & 0.12 & 38 & $-$0.12 & 0.07 & 82 & $-$0.31 & 0.09 & 20 \\
Berkeley 2 & $-$0.31 & 0.12 & 63 & $-$0.39 & 0.14 & 103 & $-$0.24 & 0.09 & 16 & $-$0.70 & 0.05 & 8 & $-$0.32 & 0.09 & 41 & $-$0.38 & 0.17 & 95 \\
Berkeley 8 & $-$0.00 & 0.17 & 414 & $-$0.41 & 0.13 & 197 & $-$0.27 & 0.05 & 36 & $-$0.29 & 0.13 & 45 & $-$0.35 & 0.06 & 229 & $-$0.21 & 0.13 & 210 \\
Berkeley 9 & 0.04 & 0.15 & 125 & $-$0.19 & 0.07 & 103 & $-$0.09 & 0.06 & 6 & $-$0.01 & 0.11 & 48 & $-$0.19 & 0.04 & 72 & $-$0.16 & 0.10 & 116 \\
Berkeley 12 & $-$0.03 & 0.13 & 239 & $-$0.35 & 0.10 & 184 & $-$0.27 & 0.06 & 33 & $-$0.30 & 0.11 & 55 & $-$0.29 & 0.05 & 122 & $-$0.26 & 0.13 & 179 \\
Berkeley 13 & $-$0.05 & 0.16 & 95 & $-$0.45 & 0.12 & 102 & $-$0.36 & 0.05 & 11 & $-$0.44 & 0.05 & 7 & $-$0.37 & 0.05 & 30 & $-$0.37 & 0.15 & 109 \\
Berkeley 14 & $-$0.14 & 0.10 & 238 & $-$0.41 & 0.16 & 131 & $-$0.29 & 0.05 & 28 & $-$0.40 & 0.07 & 54 & $-$0.37 & 0.08 & 106 & $-$0.35 & 0.19 & 133 \\
Berkeley 17 & $-$0.19 & 0.12 & 403 & $-$0.30 & 0.13 & 206 & $-$0.14 & 0.06 & 35 & $-$0.28 & 0.12 & 136 & $-$0.25 & 0.06 & 201 & $-$0.19 & 0.18 & 193 \\
Berkeley 18 & $-$0.26 & 0.12 & 582 & $-$0.38 & 0.10 & 280 & $-$0.34 & 0.06 & 86 & $-$0.59 & 0.09 & 181 & $-$0.34 & 0.06 & 248 & $-$0.38 & 0.14 & 258 \\
Berkeley 19 & $-$0.28 & 0.11 & 80 & $-$0.49 & 0.14 & 53 & $-$0.34 & 0.03 & 6 & $-$0.60 & 0.07 & 12 & $-$0.33 & 0.08 & 22 & $-$0.36 & 0.19 & 48 \\
Berkeley 21 & $-$0.20 & 0.11 & 114 & $-$0.40 & 0.14 & 65 & $-$0.33 & 0.11 & 17 & $-$0.56 & 0.11 & 47 & $-$0.37 & 0.06 & 64 & $-$0.34 & 0.19 & 65 \\
Berkeley 22 & $-$0.30 & 0.11 & 122 & $-$0.50 & 0.15 & 55 & $-$0.36 & 0.08 & 14 & $-$0.55 & 0.17 & 35 & $-$0.38 & 0.09 & 49 & $-$0.40 & 0.13 & 53 \\
Berkeley 23 & $-$0.18 & 0.12 & 118 & $-$0.48 & 0.15 & 63 & $-$0.41 & 0.05 & 14 & $-$0.34 & 0.08 & 21 & $-$0.34 & 0.06 & 31 & $-$0.30 & 0.13 & 50 \\
Berkeley 24 & $-$0.29 & 0.12 & 144 & $-$0.44 & 0.13 & 116 & $-$0.45 & 0.08 & 33 & $-$0.64 & 0.12 & 25 & $-$0.38 & 0.08 & 68 & $-$0.42 & 0.18 & 102 \\
Berkeley 25 & $-$0.36 & 0.12 & 100 & $-$0.53 & 0.15 & 43 & $-$0.43 & 0.10 & 13 & $-$0.60 & 0.09 & 29 & $-$0.38 & 0.08 & 40 & $-$0.57 & 0.20 & 40 \\
Berkeley 27 & $-$0.26 & 0.10 & 112 & $-$0.42 & 0.10 & 14 & $-$0.41 & 0.03 & 4 & $-$0.67 & 0.10 & 6 & $-$0.36 & 0.05 & 10 & $-$0.47 & 0.17 & 17 \\
Berkeley 28 & $-$0.25 & 0.10 & 30 & $-$0.58 & 0.12 & 24 & -- & -- & -- & -- & -- & -- & $-$0.42 & 0.04 & 6 & $-$0.49 & 0.18 & 21 \\
Berkeley 29 & $-$0.28 & 0.13 & 33 & $-$0.67 & 0.09 & 19 & -- & -- & -- & $-$0.57 & 0.09 & 12 & $-$0.43 & 0.08 & 19 & $-$0.61 & 0.04 & 7 \\
Berkeley 30 & $-$0.21 & 0.10 & 66 & $-$0.55 & 0.15 & 30 & $-$0.23 & 0.01 & 3 & $-$0.58 & 0.10 & 5 & $-$0.35 & 0.06 & 15 & $-$0.45 & 0.21 & 39 \\
Berkeley 31 & $-$0.29 & 0.10 & 165 & $-$0.60 & 0.16 & 47 & $-$0.40 & 0.07 & 7 & $-$0.48 & 0.14 & 23 & $-$0.39 & 0.06 & 46 & $-$0.44 & 0.22 & 38 \\
Berkeley 32 & $-$0.20 & 0.11 & 517 & $-$0.45 & 0.12 & 116 & $-$0.38 & 0.04 & 16 & $-$0.35 & 0.08 & 83 & $-$0.34 & 0.08 & 109 & $-$0.31 & 0.12 & 98 \\
Berkeley 35 & $-$0.32 & 0.11 & 141 & $-$0.56 & 0.13 & 86 & $-$0.38 & 0.09 & 13 & $-$0.40 & 0.10 & 21 & $-$0.34 & 0.06 & 26 & $-$0.42 & 0.17 & 40 \\
Berkeley 37 & $-$0.36 & 0.10 & 83 & $-$0.55 & 0.13 & 74 & $-$0.42 & 0.04 & 6 & $-$0.44 & 0.10 & 13 & $-$0.39 & 0.07 & 28 & $-$0.48 & 0.20 & 41 \\
Berkeley 39 & $-$0.25 & 0.14 & 747 & $-$0.41 & 0.15 & 374 & $-$0.20 & 0.06 & 47 & $-$0.21 & 0.10 & 210 & $-$0.28 & 0.08 & 421 & $-$0.24 & 0.13 & 311 \\
Berkeley 44 & 0.00 & 0.16 & 135 & $-$0.20 & 0.13 & 117 & 0.08 & 0.07 & 30 & 0.03 & 0.09 & 39 & $-$0.12 & 0.07 & 177 & $-$0.07 & 0.12 & 134 \\
Berkeley 52 & $-$0.36 & 0.28 & 14 & $-$0.24 & 0.10 & 53 & $-$0.24 & 0.04 & 5 & -- & -- & -- & $-$0.23 & 0.09 & 58 & $-$0.04 & 0.06 & 48 \\
Berkeley 54 & $-$0.19 & 0.12 & 125 & $-$0.20 & 0.09 & 78 & $-$0.13 & 0.11 & 11 & $-$0.23 & 0.09 & 65 & $-$0.21 & 0.08 & 88 & $-$0.13 & 0.09 & 81 \\
Berkeley 56 & $-$0.27 & 0.13 & 106 & $-$0.42 & 0.11 & 64 & $-$0.45 & 0.02 & 5 & $-$0.52 & 0.08 & 51 & $-$0.28 & 0.08 & 69 & $-$0.36 & 0.16 & 64 \\
Berkeley 60 & $-$0.30 & 0.14 & 28 & $-$0.41 & 0.15 & 82 & $-$0.18 & 0.01 & 3 & -- & -- & -- & $-$0.18 & 0.05 & 11 & $-$0.48 & 0.16 & 98 \\

... & ...& ...& ...& ...& ...& ...& ...& ...& ...& ...& ...& ...& ...& ...& ...& ...& ...& ...\\

		\hline\noalign{\smallskip}

\end{tabular}

\tablefoot{[M/H] or [Fe/H] represents the median metallicity, with $\sigma_{\text{[M/H]}}$ or $\sigma_{\text{[Fe/H]}}$ denoting its uncertainty (MAD). N indicates the number of members retrieved in each catalogue.}

\end{threeparttable}

\end{sidewaystable}

\FloatBarrier
\onecolumn
\section{\mbox{Metallicity determination for three OCs as an example}}

\begin{figure*}[!ht]
	\centering
        \includegraphics[width=56mm]{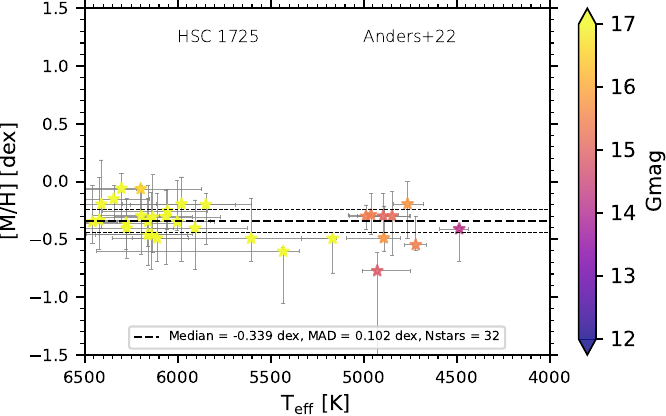}
        \includegraphics[width=56mm]{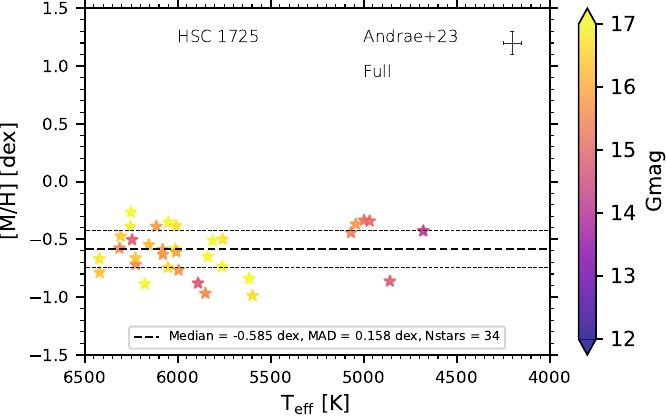}
        \includegraphics[width=56mm]{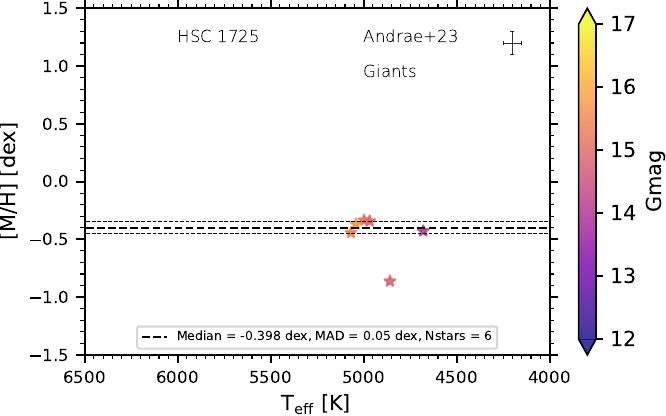}
        \includegraphics[width=56mm]{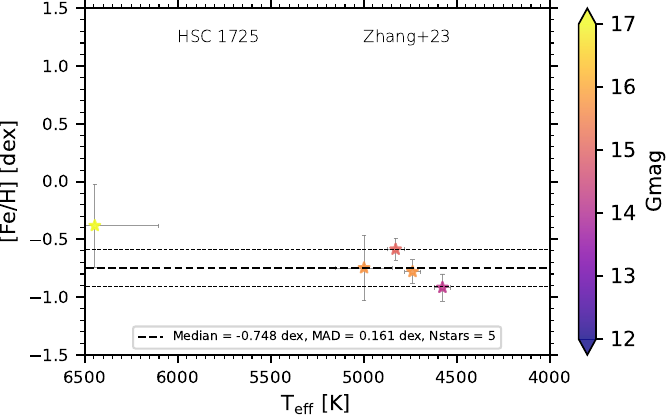}
        \includegraphics[width=56mm]{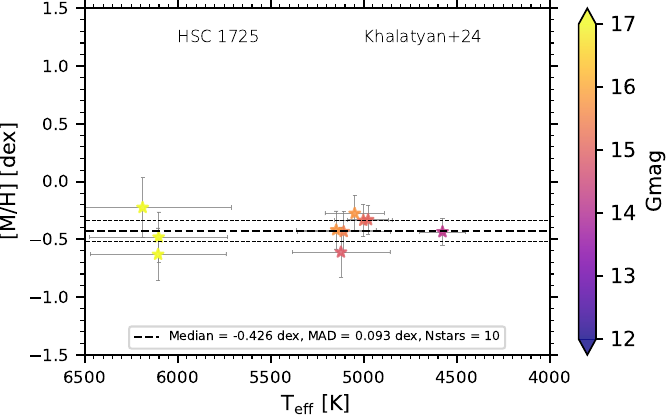}
        \includegraphics[width=56mm]{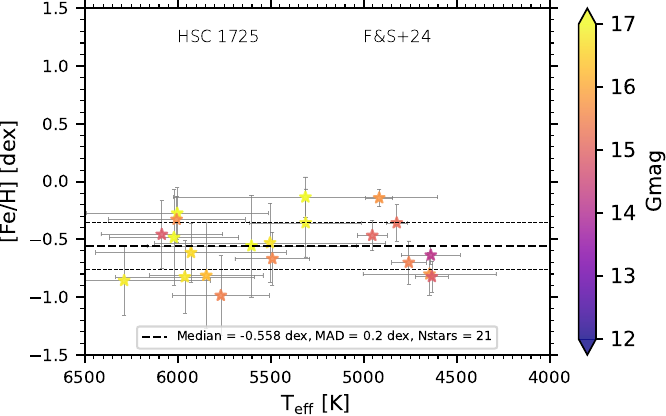}
	\caption{Metallicity [Fe/H] or [M/H] as a function of T$_{\text{eff}}$, colour-coded with G-band apparent magnitude, of cluster HSC~1725 sourced from the six catalogues.  The black dashed line in each panel marks the median metallicity of the represented members. The two thin black dashed lines represent the median plus and minus the MAD. Individual error bars are represented but for the parameters from \cite{Andrae+2023b} we represent the typical uncertainties quoted by the authors, 50 K in T$_{\text{eff}}$ and 0.1 dex in [M/H] in the top right corner of the two panels.}
	\label{fig: HSC~1725}
\end{figure*}

\begin{figure*}[!ht]
	\centering 
        \includegraphics[width=56mm]{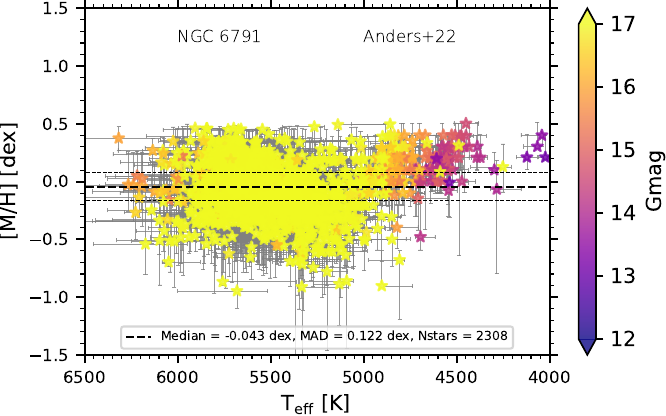}
        \includegraphics[width=56mm]{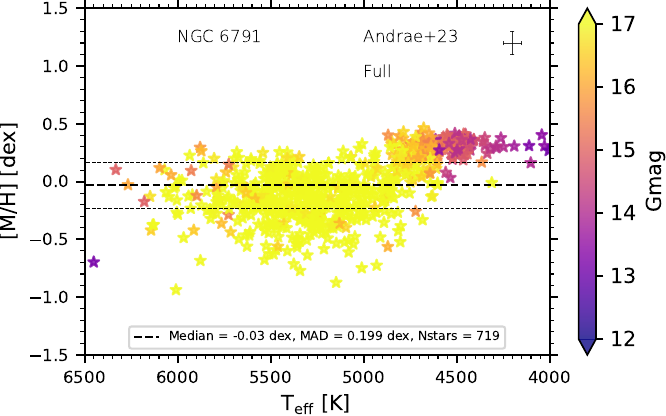}
        \includegraphics[width=56mm]{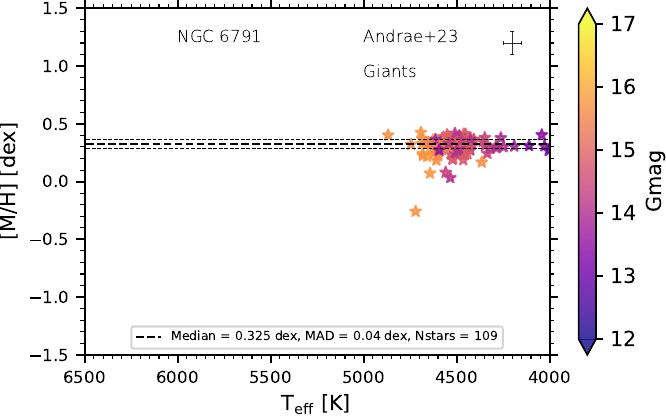}
        \includegraphics[width=56mm]{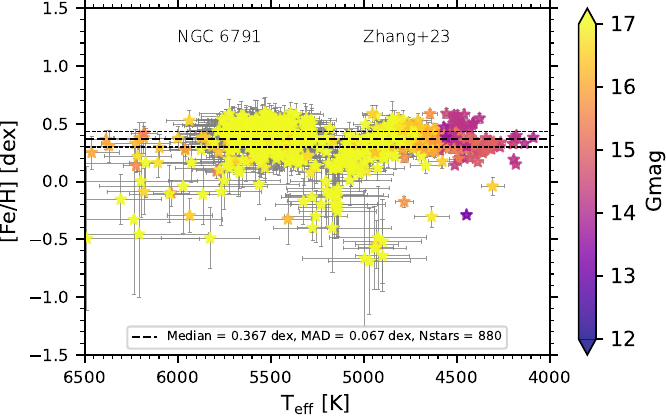}
        \includegraphics[width=56mm]{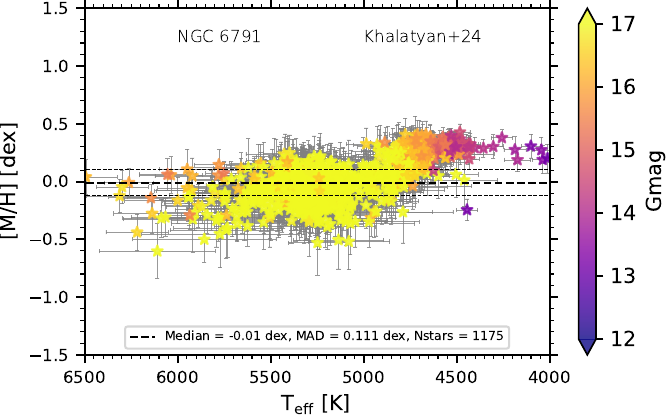}
        \includegraphics[width=56mm]{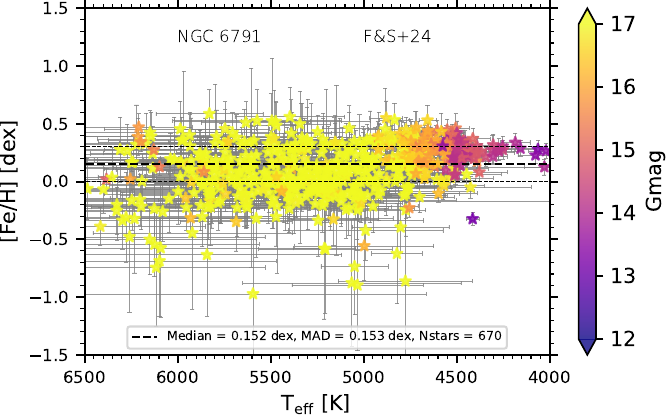}
	\caption{Same as Fig~\ref{fig: HSC~1725}, but for NGC~6791.}
	\label{fig: NGC~6791}
\end{figure*} 

\begin{figure*}[!ht]
	\centering 
        \includegraphics[width=56mm]{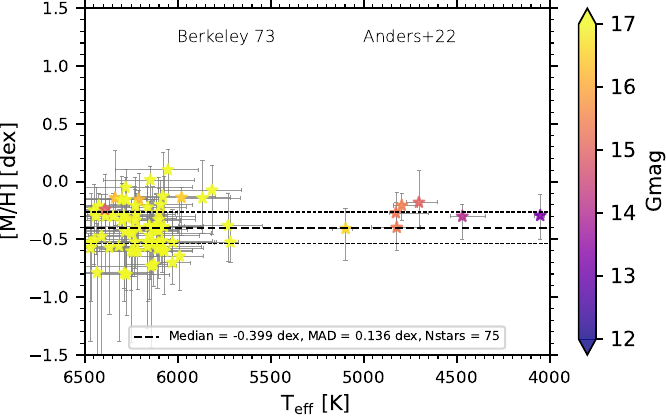}
        \includegraphics[width=56mm]{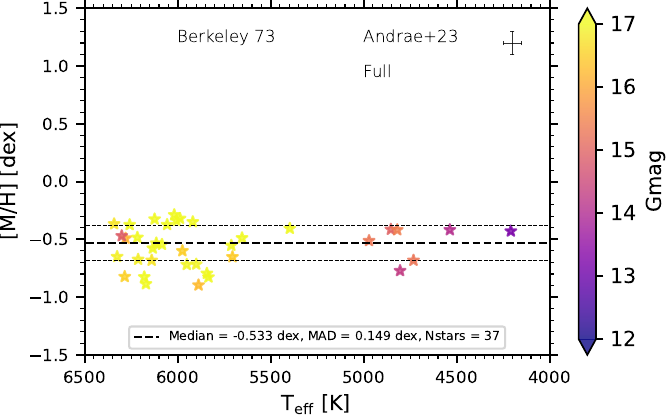}
        \includegraphics[width=56mm]{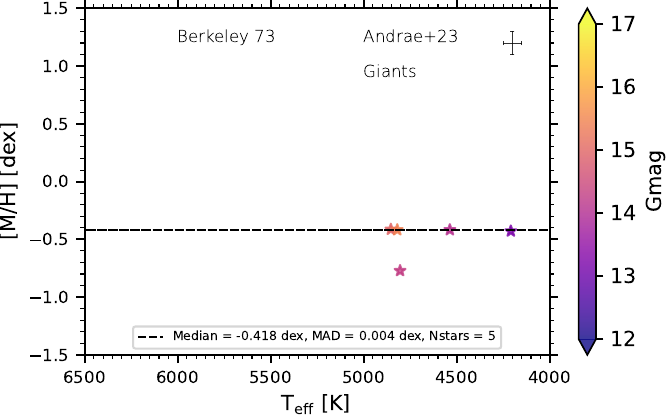}
        \includegraphics[width=56mm]{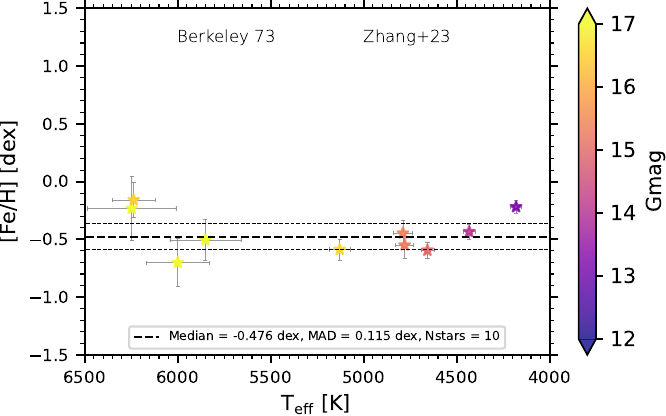}
        \includegraphics[width=56mm]{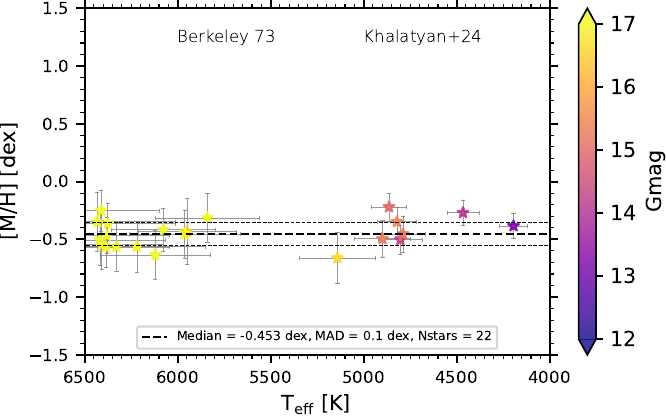}
        \includegraphics[width=56mm]{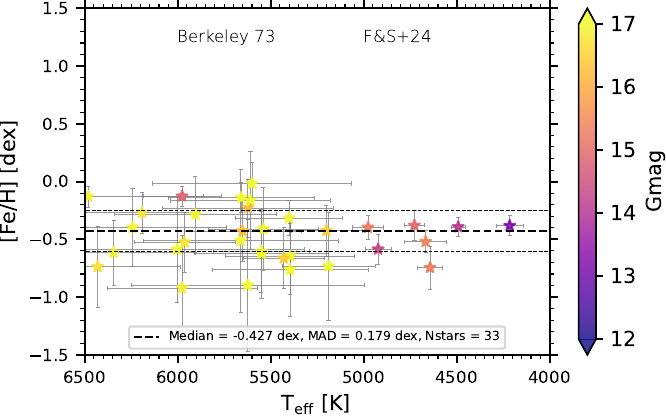}
	\caption{Same as Fig~\ref{fig: HSC~1725}, but for Berkeley~73.}
	\label{fig: Be73}
\end{figure*}

\FloatBarrier
\twocolumn

\section{Bayesian segmented linear fitting model}
\label{appendix_segmented}

Inspired by the Bayesian 'linear gradient + scatter' model in \citet{Anders+2017}, we constructed a Bayesian segmented linear fitting model in this work that actually incorporates a knee based on the Bayesian 'linear gradient + scatter' model, and applied it to the distribution between [M/H] and the galactocentric distance (R$_{\text{G}}$). It is assumed that there is a knee point in the distribution between metallicity and galactocentric distance. Therefore, we built a segmented linear function with a shared knee point, as shown in Eq.~\ref{eq:piecewise_model}, including four parameters (R$_{\text{knee}}$, b$_{\text{left}}$, m$_{\text{left}}$, m$_{\text{right}}$). R$_{\text{knee}}$ indicates the break radius, marking the critical galacocentric distance where the metallicity gradient changes, with b$_{\text{left}}$ being the intercept of the left piecewise linear function, and m$_{\text{left}}$ and m$_{\text{right}}$ denoting the slopes of the left and right piecewise linear functions, respectively. R$_{\text{G},i}$ is the galactocentric radius. This function has two properties; it is continuous at the inflection point R$_{\text{knee}}$ and the right segment intercept is determined by continuity at R$_{\text{knee}}$. A negative m$_{\text{left}}$ reflects the observed decline in metallicity with increasing galactocentric radius in the inner disc, while m$_{\text{right}}$ may be positive or negative to capture variations in the outer disc:

\begin{equation}\label{eq:piecewise_model}
\text{[M/H]}_{\text{model}}(R_\text{G},i; \theta) = 
\begin{cases} 
m_{\text{left}} R_{\text{G},i} + b_{\text{left}}, & R_{\text{G},i} < R_{\text{knee}} \\
m_{\text{right}} (R_{\text{G},i} - R_{\text{knee}}) + (m_{\text{left}} R_{\text{knee}} + b_{\text{left}}), & R_{\text{G},i} \geq R_{\text{knee}}
\end{cases}
\end{equation}

We assumed that [M/H]$_{i}$ depend linearly on the galactocentric radius, R$_{\text{G},i}$, which is convolved with a Gaussian distribution with the individual uncertainty of Gaussian measurement, $\sigma_{i}$. This allows the intrinsic [M/H] abundance spread to be linearly correlated with R$_{\text{G}}$. Due to no error in R$_{\text{G}}$, a likelihood function can be assumed to be Eq.~\ref{eq:likelihood} where  [M/H]$_{i}$ and $\sigma_{i}$ are the metallicity measurement and the uncertainty for the $i$-th star cluster. For $\sigma_{i}$, it can be assumed to have two components, i.e. the error of metallicity measurement (e$_{[M/H],i}$) and its scatter ($\sigma$). $N$ is the total number of clusters. $\theta$ thus consists of R$_{\text{knee}}$, b$_{\text{left}}$, m$_{\text{left}}$, m$_{\text{right}}$, and $\sigma$. In addition, we consider the flat distribution for the five parameters. A joint prior function, thus, is defined as the product of independent uniform priors, as shown in Eq.~\ref{eq:prior_dist}. Finally, we can list a posterior distribution (Eq.~\ref{eq:posterior}) according to Bayes' theorem, which is proportional to the product of the likelihood and prior.

\begin{equation}\label{eq:likelihood}
\mathcal{L}(\{R_{\text{G},i}, \text{[M/H]}_i\}_{i=1}^N | \theta) = \prod_{i=1}^N \frac{1}{\sqrt{2\pi \sigma_i^2}} \exp\left( -\frac{(\text{[M/H]}_i - \text{[M/H]}_{\text{model}}(R_{\text{G},i}; \theta))^2}{2\sigma_i^2} \right)
\end{equation}

\begin{equation} \label{eq:scatter}
\sigma_{i} = \sqrt{e_{[M/H],i}^2+\sigma^2}
\end{equation}

\begin{equation} \label{eq:prior_dist}
p(\theta) = p(R_{\text{knee}}) \cdot p(b_{\text{left}}) \cdot p(m_{\text{left}}) \cdot p(m_{\text{right}}) \cdot p(\sigma)
\end{equation}

\begin{equation}\label{eq:posterior}
p(\theta | \{R_{\text{G},i}, \text{[M/H]}_i\}_{i=1}^N) \propto \mathcal{L}(\{R_{\text{G},i}, \text{[M/H]}_i\}_{i=1}^N | \theta) \cdot p(\theta)
\end{equation}

Next, we employ the \verb|emcee| package with 40 Markov chains, each performing 10,000 sampling steps. By this way, we obtain the five fitting parameters of our sample in the distribution of [M/H] and R$_{\text{G}}$, as shown in the bottom left panel of Fig.~\ref{fig:Z_MFeh_relationships}.

\end{appendix}

\end{document}